\documentclass[
    a4paper,
    reprint,
    showpacs,
    amsmath,
    amssymb,
    aip,
    jcp,
    showkeys
    ]{revtex4-2}

\usepackage[utf8]{inputenc}
\usepackage{gensymb}
\usepackage{etex}
\usepackage{graphicx}
\usepackage{color}
\usepackage{colortbl}
\usepackage{xcolor}
\usepackage{calc}
\usepackage{amsfonts}
\usepackage{amsmath}
\usepackage{amssymb}
\usepackage{wasysym}
\usepackage{float}
\usepackage{epstopdf}
\usepackage{nicefrac}
\usepackage{tcolorbox}
\usepackage[makeroom]{cancel}
\usepackage{threeparttable}
\usepackage[vcentermath]{youngtab} 
\usepackage{tikz}
\usetikzlibrary{trees}
\usetikzlibrary{calc}
\usetikzlibrary{positioning}
\usetikzlibrary{shapes.misc}
\usetikzlibrary{decorations.pathreplacing}
\usetikzlibrary{shapes.arrows} 
\usetikzlibrary{backgrounds}

\tikzset{
	solid node/.style = {circle, draw, inner sep = 3, fill = black},
	right angle/.style = {grow = 300},
	left angle/.style = {grow = 240},
	short/.style = {level distance = 1cm, },
	long/.style = {level distance = 2cm},
	left up/.style = {grow = 120},
	right up/.style = {grow = 60}
}

\usepackage{rotating}
\usepackage{mathtools}
\usepackage{units}
\usepackage{tabularx}
\usepackage{pgf}
\usepackage{pgfplots}
\usepackage{geometry}
\usetikzlibrary{calc,intersections}
\usepackage{relsize}
\usepackage{longtable}
\usepackage{dcolumn}
\usepackage{braket}
\usepackage{listings}
\usepackage{algorithm}
\usepackage{relsize}
\usepackage{nicefrac}
\usepackage{diagbox}
\usepackage{standalone}
\usepackage{bm}
\usepackage{multirow}
\usepackage{acronym}
\usepackage{tikzorbital}
\usepackage{setspace}
\usepackage{bbold}
\usepackage[inline]{enumitem}
\pgfplotsset{compat=newest}
\usepackage{geometry}
\usepackage[version=4]{mhchem}
\usepackage{booktabs}
\usepackage{xr-hyper}
\usepackage{hyperref}

\usepackage{colortbl}

\newcommand{\eqnref}[1]{~(\ref{#1})}%

	{%
	\end{list}%
}%

\newlength{\marginwidth}
\providecommand{\abs}[1]{\lvert#1\rvert}

\newcommand{\der}{\partial\mspace{2mu}} 

\DeclareMathOperator{\e}{e} 			

\newcommand\commutator[2]{\ensuremath{\mathinner{%
			\mathopen[\,#1,#2\,\mathclose]}}}

\newcommand*{\dashfill}{\leavevmode\cleaders\hbox{-}\hfill\kern0pt}

\newcommand*{\midhrulefill}{
	\leavevmode
	\cleaders\hbox to 1ex{\raisebox{.5ex}{\rule{1ex}{.4pt}}}\hfill\kern0pt
}

\newcommand*{\braopket}[3]{\ensuremath{\langle{#1}|{#2}|{#3}\rangle}}

\renewcommand{\d}{\downarrow}
\renewcommand{\u}{\uparrow}
\newcommand{\s}{\sigma}

\newcommand{\mbf}[1]{\mathbf{#1}}

\newcommand{\neci}{\texttt{NECI}}

\newcolumntype{L}{>{$}l<{$}} 
\newcolumntype{C}{>{$}c<{$}} 

\newcommand{\ra}{\ensuremath{\rightarrow}}

\newcommand*{\citen}{}
\DeclareRobustCommand*{\citen}[1]{%
	\begingroup
	\romannumeral-`\x 
	\setcitestyle{numbers}%
	\cite{#1}%
	\endgroup
}

\newcommand{\cre}[1]{a_{#1}^\dagger}
\newcommand{\ann}[1]{a_{#1}^{\phantom{\dagger}}}
\newcommand{\num}[1]{n_{#1}}

\newcommand{\etal}{\emph{et al.}}

\definecolor{spawncolor}{RGB}{200 0 0}
\definecolor{deathcolor}{RGB}{0 200 0}
\definecolor{annihilcolor}{RGB}{255 99 71}
\definecolor{endcolor}{RGB}{255 99 71}

\tikzset{state node/.style={circle,thick,inner sep=1pt}}

\tikzset{state conn/.style={very thick}}
\tikzset{state dash/.style={very thick, dashed}}

\tikzset{walker/.style={very thick}}
\tikzset{walker new/.style={very thick, red}}
\tikzset{spawn/.style={very thick, color=spawncolor}}
\tikzset{death/.style={very thick, color=deathcolor}}
\tikzset{annihil/.style={very thick, color=annihilcolor}}
\newlength{\shortlength}
\newlength{\longlength}
\newlength{\afourwidth}
\newlength{\jctcsingle}
\newlength{\jctcdouble}
\newlength{\walkerlength}
\newlength{\tablesep}
\newlength{\gridsep}
\tikzset{branch/.style={circle,draw,thick,inner sep=1pt}}
\tikzset{growd/.style={-,very thick, red}}
\tikzset{grid/.style={-}}

\tikzset{lattline/.style={-,thick}}

\newlength{\orbsep}
\newlength{\sepone}
\newlength{\septwo}
\newlength{\septhree}
\tikzset{graph/.style={circle,draw,thick,inner sep=2pt}}
\tikzset{graphline/.style={-,thick}}

\tikzset{site/.style={circle,very thick,draw,opacity=0.8,inner sep=1pt}}

\newlength{\latticesep}
\newlength{\indicator}
\tikzset{orbline/.style={-,very thick}}

\tikzset{doubspin/.style={-,very thick}}
\tikzset{posspin/.style={-,very thick,blue}}
\tikzset{negspin/.style={-,very thick,red}}

\newlength{\picsep}
\newlength{\stepzero}
\tikzset{loop-tail/.style={circle,thick,inner sep=1pt,draw,fill}}
\tikzset{tree-out/.style={circle,thick,inner sep=1pt,draw,fill}}

\tikzset{mline/.style={thick}}

\newlength{\onedist}
\newlength{\twodist}
\newlength{\threedist}
\tikzset{inter/.style={circle,thick,inner sep=1pt,draw}}
\tikzset{mprime/.style={thick,dashed}}

\newlength{\taildist}
\newlength{\headdist}
\tikzset{line-out/.style={thick,dotted}}

\newlength{\gridlen}
\tikzset{drt-vert/.style={circle,very thick,draw}}

\tikzset{tree-start/.style={rectangle,very thick,draw,rounded corners=2pt,text width = {width("1")}, text height = {height("1")}, outer sep = 0,inner sep = 4pt}}
\tikzset{tree-mid/.style={circle,very thick, draw,  text width = {width("1")}, text height = {height("1")}, outer sep = 0, inner sep = 2pt,align=center}}

\colorlet{csf-color-2}{orange!80}
\colorlet{csf-color-1}{blue!60}

\tikzset{fixed-node/.style={circle,very thick, draw,  text width = {width("10")}, text height = {height("10")}, outer sep = 0, inner sep = 1pt,align=center}}

\tikzset{empty-node/.style={circle,very thick,  text width = {width("10")}, text height = {height("10")}, outer sep = 0, inner sep = 1pt,align=center}}

\definecolor{doublecolor}{RGB}{200 0 0}
\definecolor{singlecolor}{RGB}{0 200 0}

\tikzset{lattline/.style={-,thick}}
\tikzset{fermi/.style={rounded corners}}
\tikzset{line/.style={very thick}}
\tikzset{arrow/.style={->,very thick}}
\tikzset{box/.style={fill=white,rounded corners=1pt, opacity = 0.95, text opacity = 1.0}}

\tikzset{gridline/.style={dashed,opacity = 0.2}, very thin}

\definecolor{navyblue}{RGB}{0 0 128}
\definecolor{cadetblue}{RGB}{95 158 160}
\definecolor{branchingcolor}{RGB}{210 105 30}

\definecolor{back_fill}{gray}{0.9}
\definecolor{back_fill_dark}{gray}{0.7}

\tikzset{end-node/.style = {tree-out, ,text width = 2pt,text height = 2pt}}

\tikzset{l-long/.style= {<-,thick,out = 315, in = 45}}
\tikzset{r-long/.style= {->,thick,out = 315, in = 45}}

\tikzset{l-short/.style= {<-,thick,out = 315, in = 45}}
\tikzset{r-short/.style= {->,thick,out = 315, in = 45}}
\tikzset{bubble/.style= {->,thick,out = 340, in = 20}}
\tikzset{bubble-left/.style= {->,thick,out = 225, in = 110}}

\tikzset{r-long-left/.style= {->,thick,out = 225, in = 135}}
\tikzset{l-long-left/.style= {<-,thick,out = 225, in = 135}}
\tikzset{r-short-left/.style= {->,thick,out = 225, in = 135}}
\tikzset{l-short-left/.style= {<-,thick,out = 225, in = 135}}

\pgfdeclarelayer{background}
\pgfdeclarelayer{foreground}
\pgfsetlayers{background,main,foreground}

\tikzset{flow-box/.style={rectangle,very thick,draw,rounded corners=2pt}}

\tikzset{flow-box-end/.style={rectangle,very thick,draw,rounded corners=2pt}, draw = red}
\tikzset{flow-arrow/.style={->,very thick,rounded corners}}

\newcommand{\sref}[1]{\hyperref[{#1}]{SI.\ref*{#1}}}
\begin{document}

\author{Werner Dobrautz}
\email{w.dobrautz@fkf.mpg.de}
\affiliation{Max Planck Institute for Solid State Research, Heisenbergstr. 1, 70569 Stuttgart, Germany}%

\author{Oskar Weser}
\affiliation{Max Planck Institute for Solid State Research, Heisenbergstr. 1, 70569 Stuttgart, Germany}%

\author{Nikolay Bogdanov}
\affiliation{Max Planck Institute for Solid State Research, Heisenbergstr. 1, 70569 Stuttgart, Germany}%

\author{Ali Alavi}
\affiliation{Max Planck Institute for Solid State Research, Heisenbergstr. 1, 70569 Stuttgart, Germany}%
\affiliation{Yusuf Hamied Department of Chemistry, University of Cambridge, Lensfield Road, Cambridge CB2 1EW, United Kingdom}%

\author{Giovanni Li Manni}
\affiliation{Max Planck Institute for Solid State Research, Heisenbergstr. 1, 70569 Stuttgart, Germany}%

\title{Spin-pure Stochastic-CASSCF via GUGA-FCIQMC applied to Iron Sulfur Clusters}

\date{\today}

%



\begin{abstract}

In this work we demonstrate how to compute the one- and
two-body reduced density matrices within the
\emph{spin-adapted} full configuration interaction quantum 
Monte Carlo (FCIQMC) method, which is based
on the graphical unitary group approach (GUGA).
This allows us to use GUGA-FCIQMC as a spin-pure
configuration interaction (CI) eigensolver within 
the complete active space self-consistent field 
(CASSCF) procedure, and hence to stochastically
treat active spaces far larger than conventional CI
solvers whilst variationally relaxing orbitals 
for specific spin-pure states. We apply the method to investigate  
the spin-ladder in iron-sulfur dimer and tetramer model systems.
We demonstrate the importance of the orbital relaxation by 
comparing the Heisenberg model magnetic coupling parameters from  
the CASSCF procedure to those from a CI-only procedure based on restricted open-shell 
Hartree-Fock orbitals.
We show that orbital relaxation differentially stabilizes the lower
spin states, thus enlarging the coupling parameters with respect
to the values predicted by ignoring orbital relaxation effects.
Moreover, we find that while CI eigenvalues are well fit
by a simple bilinear Heisenberg Hamiltonian, the CASSCF 
eigenvalues exhibits deviations that necessitate the inclusion of 
biquadratic terms in the model Hamiltonian.

\end{abstract}

\maketitle
\section{\label{sec:intro}Introduction}

The complete active space self-consistent field (CASSCF) method is
a well-established approach in quantum chemistry for the treatment 
of strongly correlated electron systems with substantial multi-reference 
character.\cite{Sigbahn1980,Siegbahn1981,Helgaker2000,Roos1980,Roos1987,LiManni2016,Dobrautz2020, Ruedenberg1976}
Important static correlation effects are rigorously described within the \emph{active space},
consisting of the most important orbitals and electrons,
while the effect of the \emph{environment} (electrons not included in the active space) is
accounted for at the mean-field level via a variational orbital optimization 
(the SCF procedure).
One- and two-body reduced density matrices (1- and 2-RDMs)
within the active space are necessary to perform orbital rotations
between the active orbitals and the environment, whether a second-order 
Newton-Raphson formulation,~\cite{Siegbahn1981, Kreplin2019, Knowles1985, Werner1985,Dobrautz2020}  or the simplified 
Super-CI technique with an average Fock operator is utilized.~\cite{Roos1980}
If applicable, exact diagonalization techniques~\cite{lanczos1950, Sleijpen2000, Davidson1975}
are utilized to obtain eigenvalues, eigenvectors and the RDMs 
associated to the CAS configuration interaction (CASCI) Hamiltonian.
However, due to the exponential scaling of CASCI with respect to the size of the active space, 
exact diagonalization techniques are restricted to at most about 18 electrons in 18 orbitals, CAS(18e,18o),
on serial architecture.~\cite{Molcas2016, OpenMolcas2019}
More recent massively parallel implementations allow sizes up to CAS(24e,24o).~\cite{Vogiatzis2017}
Another strategy is to use methods that approximate the full-CI wave function in the active space, 
like the density matrix renormalization group approach
(DMRG),~\cite{White1992, White1993, Schollwock2011, Schollwock2005,  Nakatani2017, Zgid2008, Chan2008, Veis2021, Sharma2011, Chan2012, Reiher2010, Reiher2015, Reiher2017} 
full configuration interaction quantum Monte Carlo (FCIQMC),~\cite{Guther2020, linear-scaling-fciqmc, Alavi2009, Alavi2010, LiManni2016,Thomas2015}
selected configuration interaction (Selected-CI) approaches,~\cite{Rancurel1973,
	Malrieu1983, Caffarel2017, Levine2020, Smith2017, Sharma2017a, Holmes2016b, Tubman2020, Tubman2016, Garniron2018}
recently implemented in a spin-adapted form\cite{Zhang2020, Chilkuri2021, Chilkuri2021b}, 
as well as the generalized active space approach,~\cite{Ma2011,Vogiatzis2015,Weser2021} as
CI-eigensolvers within the CASSCF framework.
These approaches allow the study of much larger active spaces.~\cite{LiManni2016,LiManni2018,Weser2021,Bogdanov2018,LiManni2019, Levine2020, Sun2017, Sharma2017b,LiManni2020,LiManni2020b,LiManni2021}
The use of FCIQMC as the CASSCF CI-eigensolver within the Super-CI framework, 
termed Stochastic-CASSCF,~\cite{LiManni2016} has been developed in our group, 
and used to study a number of strongly correlated systems,
such as model systems of Fe(II)-porphyrins and the correlation 
mechanisms that differentially stabilize the intermediate 
spin states over the high spin states,~\cite{LiManni2018,LiManni2019,Weser2021}
and model systems of corner-sharing cuprates.~\cite{Bogdanov2018}
The original Stochastic-CASSCF implementation was formulated 
using Slater-Determinants (SDs) as many-body basis for 
the CASCI wave function expansion.
As SDs are not necessarily eigenfunctions of the total spin operator,
its applicability is bound to
the intrinsic  spin structure of the system studied.
If the low-spin states are energetically more stable and well separated 
from higher spin states, it is possible to obtain essentially spin-pure wave functions
when using an SD basis.
However, when high-spin states are more stable than low-spin states, 
and/or a number of spin states are nearly degenerate, 
it is very difficult to obtain spin-pure solutions,
or target states other than the ground state, with a SD basis.

In this paper we present an algorithm for the calculation
of 1- and 2-RDMs within the spin-adapted implementation of FCIQMC
via the graphical unitary group approach (GUGA-FCIQMC).~\cite{Dobrautz2019}
GUGA-FCIQMC has been implemented within the \texttt{NECI} code,~\cite{Guther2020, neci} 
and  provides accurate spin-adapted wave functions and RDMs for
active space sizes out-of-reach for conventional exact
CI-eigensolvers.~\cite{LiManni2020,LiManni2020b}
As already done for the original Stochastic-CASSCF~\cite{LiManni2016},
the sampled 1- and 2-RDMs are then utilized within the Super-CI
procedure as implemented in the \texttt{OpenMolcas} chemistry software
package,~\cite{OpenMolcas2019} to perform the orbital relaxation step.
Thus, via the interface of the NECI code and OpenMolcas~\cite{OpenMolcas2019},
it is possible to perform spin-adapted \emph{state-specific}
(or \emph{state-average}, if RDMs of different states
are weighted-averaged prior the Super-CI step) Stochastic-CASSCF
optimizations, targeting any desired spin state.
The spin-pure Stochastic-CASSCF allows us to 
obtain variationally optimized molecular orbitals,
which in turn enable the calculation of spin gaps, 
unbiased from the choice of the starting orbitals.

The applicability and the importance of the method is shown through the investigation
of the spin ladder of two iron-sulfur (FeS) clusters.
Poly-nuclear transition-metal (PNTM) clusters are of major importance in organometallic chemistry
and as cofactors in biology, and are involved in a multitude of processes,
including photosynthesis,
respiration and nitrogen fixation \cite{Beinert1997, Osterberg1974, Howard1996},
being responsible for redox reactions~\cite{Vollmer1983, Stombaugh1976, Rees2003}
and electron transfer,~\cite{Hudson2005, Mortenson1962, Tagawa1962, Lubitz2014, 
	Blondin1990, Han1989, Mitchell1985, Golbeck1987, Peters1997}
act as catalytic agents and even provide a redox sensory function.~\cite{Ibrahim2020}
A theoretical understanding of the intricate interplay of the energetically low-lying spin states 
of these systems, guided by accurate numerical results, could provide insights towards
the synthetic realization of these processes.
Especially, because direct experimental measurements targeting the electronic 
structures of these systems are often hindered by the large
number of overlapping electronic states, and corresponding 
vibrational modes at finite temperatures.~\cite{Johnson2005,Noodleman1995,DeBeer2018}
In addition, some energetically low-lying excited states are inaccessible
by accurate optical absorption experiments due to being electric-dipole 
forbidden transitions.~\cite{Eaton1971}

Spin-pure stochastic RDM sampling allows us to formulate a spin-adapted Stochastic-CASSCF and gives 
us access to properties encoded in the 1- and 2-RDMs, such as spin-spin correlation functions.
Using CASSCF wave functions of various active space size and composition we will study and discuss 
how spin gaps are affected by orbital relaxation effects.
Additionally, the \emph{ab initio} energies will be mapped to the (biquadratic) Heisenberg 
spin model\cite{Heisenberg1928,Dirac1926,Boca1999,Malrieu2008,Kittel1960,Anderson1959, 
	Falk1984,Bastardis2008,Calzado1998,Bastardis2007,Moreira2002} to show the effect of 
active space size and orbital relaxation on the extracted magnetic coupling parameters,
which are in turn compared to the available experimental data~\cite{Gillum1976,Palmer1971} 
and other computational studies.~\cite{Sharma2014,Noodleman1992}

The remainder of this paper is organized as follows:
In Sec.~\ref{sec:guga-fciqmc} we summarize the spin-adapted GUGA-FCIQMC method and
in Sec.~\ref{sec:guga-rdms} we describe the sampling algorithm of spin-free RDMs.
In Sec.~\ref{sec:results} we discuss \emph{ab initio} CASSCF spin gaps and spin-spin
correlation functions for an iron-sulfur dimer, $\ce{Fe_{2}S_{2}}$,~\cite{LiManni2020}
for different active space sizes and starting orbitals, 
and for an $\ce{[Fe_{4}S_{4}]}$ tetramer model system.
We also map our \emph{ab initio} results to a (biquadratic) Heisenberg model Hamiltonian, 
discuss the role of the CASSCF procedure when extracting the exchange 
parameter(s), and compare the magnetic coupling constants extracted from our computations
to experimental and theoretical references.
Finally, in Section~\ref{sec:conclussion} we summarize 
our findings and offer a general discussion on the presented topic.

An appendix is available, where we derive necessary
formulas for local spin measurements (Appendix~\ref{sec:local-spin}),
and spin-spin correlation functions from RDMs (App.~\ref{app:spin-corr}).
We additionally supply coordinate and orbital files,
computational details, and comparisons with available exact results for small active spaces, a table with the data used in Fig.~\ref{fig:singlet-22in26-wf}, a study on improved convergence due to stochastic noise, 
the protocol how we compared the orbitals in Fig.~\ref{fig:orbial-difference-rohf}, 
details on interface and the RDM storage convention in \texttt{OpenMolcas} and a quick access literature overview of computation results for the $\ce{Fe_{2}S_{2}}$ system in the supporting information (SI).\cite{SI}

\section{\label{sec:guga-fciqmc}GUGA-FCIQMC}
In this section we  briefly summarize the main details of
the GUGA-FCIQMC implementation. More theoretical and technical 
aspects of the algorithm are available in the
literature.~\cite{dobrautz-phd,Dobrautz2019}

The spin-adapted implementation of the FCIQMC algorithm relies on the
unitary group approach (UGA),~\cite{Paldus1974,Paldus2020} pioneered
by Paldus, and its graphical extension (GUGA), introduced by
Shavitt.~\cite{Shavitt1977,Shavitt1978}
GUGA provides an efficient-to-use spin-adapted many-body basis,
based on the spin-free formulation of quantum chemistry.~\cite{Matsen1964}
The spin-free form of the electronic Hamiltonian is given by
\begin{equation}
\label{eq:spin-free-h}
\hat H = \sum_{ij}^n t_{ij} \hat E_{ij} + \frac{1}{2}\sum_{ijkl}^n V_{ijkl} \hat e_{ij, kl},
\end{equation}
with the spin-free excitation operators,
$\hat E_{ij} = \sum_\sigma \hat a_{i\sigma}^\dagger \hat a_{j\sigma}$ and 
$\hat e_{ij, kl} = \hat E_{ij}\hat E_{kl} - \delta_{jk} \hat E_{il}$,
defined in terms of the creation and annihilation operators
$\hat a_{i\sigma}^\dagger, \hat a_{j\sigma}$ with spatial orbitals 
$i, j$ and spin $\sigma$. 
$t_{ij}$ and $V_{ijkl}$ represent the one- and two-electron 
integrals in a molecular orbital basis, and $n$ indicates
the total number of spatial orbitals.

The name \emph{unitary group} approach comes from the fact that the
operators $\hat E_{ij}$ fulfill the same commutation relations as 
the generators of the unitary group of 
order $n$, $U(n)$.~\cite{Paldus1974}
Paldus~\cite{Paldus1975, Paldus1976} identified a very efficient construction of a spin-adapted 
basis tailored for the electronic structure problem, based on 
the Gel'fand-Tsetlin basis,~\cite{gelfand-1, gelfand-2, gelfand-3}
a general basis for any unitary group $U(n)$.
The graphical extension, GUGA, provides 
an efficient way to calculate Hamiltonian matrix elements, 
$\braopket{\nu}{\hat H}{\mu}$ between different CSFs, 
$\ket{\mu}$ and  $\ket{\nu}$, within a chosen spin-symmetry sector.
The combination of an efficient protocol for computing Hamiltonian 
matrix elements, and storage of the CI coefficients, enables
an effective spin-adapted formulation of exact CI eigensolvers, 
such as CAS~\cite{Sigbahn1980} and GAS~\cite{Ma2011}, perturbation theory methodologies, 
such as CASPT2~\cite{caspt2-2}, GASPT2~\cite{Ma2016} and SplitGAS~\cite{LiManni2013} as well as
the FCIQMC approach within the GUGA framework.~\cite{Dobrautz2019}

The FCIQMC algorithm~\cite{Alavi2009,Alavi2010} is based on the 
imaginary-time ($\tau = \text{i}t$) Schr{\"o}dinger equation,
\begin{equation}
\label{eq:imag-schrodinger}
\frac{\der \ket{\Psi(\tau)}}{\der \tau} = - \hat H \ket{\Psi(\tau)} \quad \stackrel{\int \text{d} \tau }{\rightarrow} \quad \ket{\Psi(\tau)} = \e^{- \tau \hat H}\ket{\Phi(0)}
\end{equation}
which, after formal integration and a first-order Taylor expansion yields 
an iterable expression for the eigenstate, $\ket{\Psi(\tau)}$
\begin{equation}
\label{eq:first-order-approx}
\Psi(\tau+\Delta \tau) \approx \left(1 - \Delta \tau \hat H \right) \Psi(\tau).
\end{equation}
FCIQMC stochastically samples the FCI wave function, $\ket{\Psi(\tau)}$, 
of a system by a set of so-called \emph{walkers} and yields estimates for the
ground- and excited-state~\cite{Blunt2015-excited} energies and properties~\cite{Blunt2017} 
via the one- and two-body RDMs.~\cite{Alavi2014}
Theoretical and algorithmic details on FCIQMC can be found in the
literature,~\cite{Alavi2009, Alavi2010} especially in the recently published review article~[\citen{Guther2020}].

At the heart of the FCIQMC algorithm is the so called \emph{spawning step}, 
which stochastically samples the off-diagonal contribution to 
the imaginary-time evolution of the targeted state,
\begin{equation}
\label{eq:spawning-step}
c_{\mu}(\tau + \Delta \tau)  = -\Delta \tau \sum_{\nu\neq \mu} H_{\mu \nu} c_\nu(\tau)
\approx -\Delta \tau \frac{H_{\mu \nu} c_\nu(\tau)}{p_{\text{gen}}(\mu\vert \nu)},
\end{equation}
with $c_\nu(\tau)$ being the coefficient of basis state function $\ket{\nu}$ 
at the imaginary-time $\tau$, of the FCI expansion 
$\ket{\Psi(\tau)} = \sum_\nu c_\nu(\tau) \ket{\nu}$ and
$p_{\text{gen}}(\mu\vert \nu)$ is the so-called \emph{generation probability} 
of choosing configuration $\ket{\mu}$ given $\ket{\nu}$.

During a FCIQMC simulation only coefficients that are at least occupied 
by a chosen minimum number of walkers (usually set to be the real number 1.) 
are kept in memory.
The off-diagonal contribution in Eq.~\eqref{eq:spawning-step} is 
then approximated by allowing each walker on each occupied configuration 
$\ket{\nu}$ to \emph{spawn} new walkers on configuration $\ket{\mu}$ 
with a non-zero Hamiltonian matrix element $\braopket{\mu}{\hat H}{\nu}$.
The process of suggesting a new configuration $\ket{\mu}$ given $\ket{\nu}$,
called the \emph{excitation generation step}, is of utmost importance.

\begin{figure}
	\includegraphics[width=0.4\textwidth]{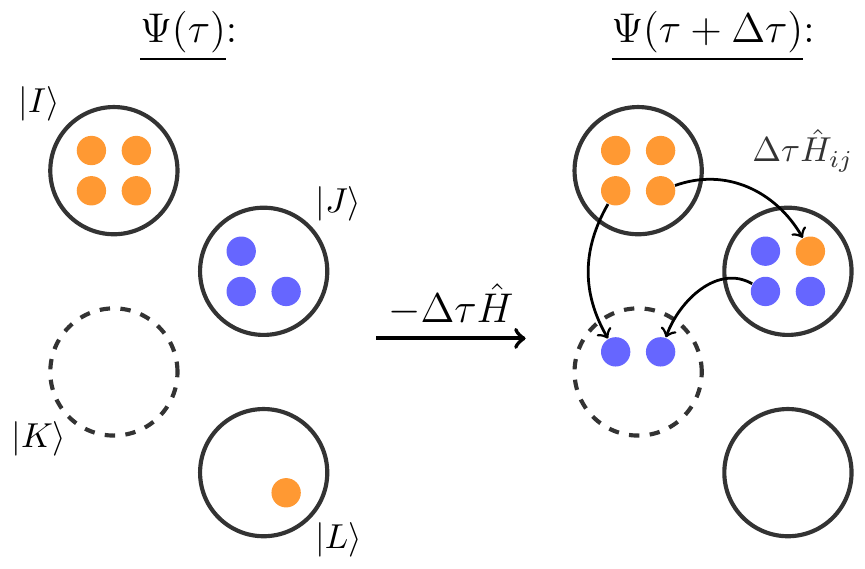}
	\caption{\label{fig:spawning}Schematic presentation of the FCIQMC spawning step. Orange and blue dots indicate opposite signed walkers on the stored basis states (black  circles). Not stored states within a time-slice are indicated by dashed circles. The arrows point towards the newly spawned children after time $\Delta\tau$ has elapsed.}
\end{figure}

The maximal usable time-step of the simulation is limited by the
relation
\begin{equation}
\label{eq:time-step}
\Delta \tau \frac{\abs{H_{\mu\nu}}}{p_{\text{gen}}(\mu\vert \nu)} \approx 1,
\end{equation}
to ensure stable dynamics.
Hence, for large $\abs{H_{\mu\nu}}/p_{\text{gen}}(\mu\vert \nu)$ ratios, the time-step of the
calculation, $\Delta \tau$, has to be lowered to ensure a stable simulation
and is the motivation for optimizing the \emph{excitation generation} step.
Several schemes to obtain a close-to-optimal balance of computational effort and
matrix element relation have been developed, see References~[\citen{Holmes2016a, Neufeld2018, Guther2020, cauchy-schwarz-1, Weser2021}].
The spawning step is schematically shown in Fig.~\ref{fig:spawning}.

\begin{figure}
	\includegraphics[width=0.35\textwidth]{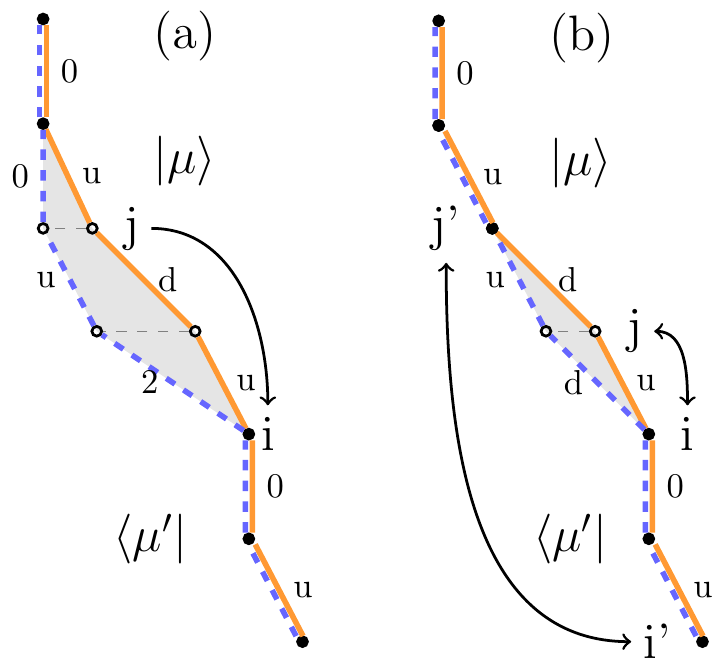}
	\caption{\label{fig:exchange-example}(a) Graphical representation of a possible single excitation from CSF $\ket{\mu} = \ket{u,0,u,d,u,0}$ to $\ket{\mu'} = \ket{u,0,2,u,0,0}$ by moving an electron from orbital $j = 5$ to $i = 3$ (indicated by the arrow on the left). The loop contributing to the coupling coefficient, $\braopket{\mu}{\hat E_{ij}}{\mu'}$, is indicated by the gray area. Following Shavitt's convention the CSFs are drawn from bottom to top. 
		(b) Exchange excitation example, for the same CSF $\ket{\mu}$, which shows that different index combinations for exchange excitations, $\hat e_{ij, ji}$ and $\hat e_{i'j', j'i'}$, can lead to the same transition $\ket{\mu} \ra \ket{\mu'} = \ket{u,0,d,u,u,0}$, with a non-zero coupling coefficient.}
\end{figure}

For reasons of interpretability, control, and improved convergence properties,
a spin-adapted implementation of FCIQMC was long-sought after.~\cite{Smart2013}
GUGA allows an efficient spin-adapted FCIQMC implementation,
by constructing spin-symmetry allowed excitations as stochastic walks 
on the graphical representation of CSFs, the so-called Shavitt graph, 
as depicted in Fig.~\ref{fig:exchange-example},
and explained in more detail in References~[\citen{Dobrautz2019, dobrautz-phd}].

\begin{figure}
	\includegraphics[width=0.3\textwidth]{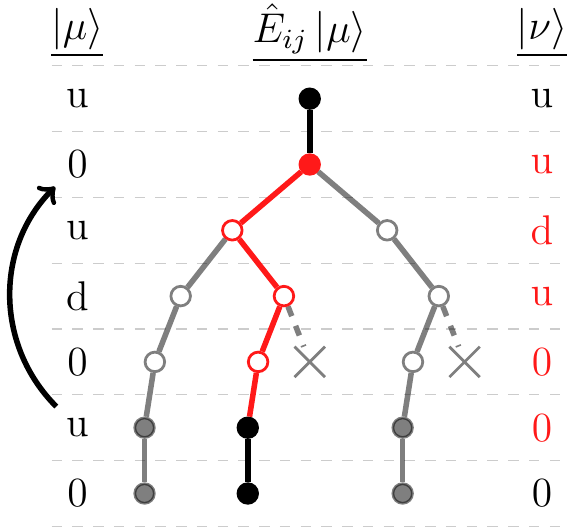}
	\caption{\label{fig:branching-tree}Schematic representation of the branching-tree approach 
		to allow efficient on-the-fly excitation generation and matrix element calculation entirely in the space of CSFs without any reference to SDs. Example given for a single excitation $\hat E_{ij}$ from a CSF $\ket \mu = \ket{u,0,u,d,0,u,0}$ to one other $\ket{\nu} = \ket{u,u,d,u,0,0,0}$.
		In the shown example, an electron is excited from the singly occupied orbital 6 in $\ket \mu$ to the empty orbital 2 
		(indicated by the arrow on the left). Spin-allowed excitation pathways are indicated by solid lines. 
		During the random excitation process in GUGA-FCIQMC, a spin-symmetry allowed path is chosen at random, weighted according to the resulting coupling coefficient, $\braopket{\mu}{E_{ij}}{\nu}$ (indicated by the red pathway). In general, empty starting orbitals and singly occupied orbitals in the excitation range allow for two possible spin couplings ($u/d$). Spin-symmetry-forbidden paths are indicated by the crossed-out nodes.
	}
\end{figure}

The GUGA allows both an efficient on-the-fly matrix element calculation and a way 
to select excitations from CSF $\ket \mu \ra \ket \nu$, and ensures the approximate relation $p_{\text{gen}}(\nu\vert \mu) \propto \abs{H_{\mu\nu}}$, via a so-called \emph{branching tree} approach. 
The stochastic GUGA excitation process for a single excitation, $\hat E_{ij}$, is schematically depicted in Fig.~\ref{fig:branching-tree}, with the CSFs drawn from top to bottom.
For a given CSF, $\ket \mu$, and two spatial orbitals, $i$ and $j$, which are 
chosen with a probability weighted according to the magnitude of their \emph{integral} contributions, at each open-shell orbital $k$ within the range $i \ra j$, an allowed 
path is chosen randomly. This process is weighted with the so-called \emph{probabilistic weight}, of the remaining decision tree below the current orbital $k$, which 
ensures the desired relation $p_{\text{gen}}(\nu\vert \mu) \propto \abs{H_{\mu\nu}}$. 
Interested readers are referred to Refs.~[\citen{Dobrautz2019, dobrautz-phd}] for more details. 

This stochastic process, additionally circumvents the bottleneck given by the exponentially
growing connectivity between CSFs with respect to the number of open-shell orbitals.
Hence, our GUGA-FCIQMC method  is able to treat systems with more than
30 open-shell orbitals and due to the spin-pure formulation, it allows to specifically target any spin-symmetry sector,
removes any spin-contamination, reduces the Hilbert space size, and
speeds up convergence in systems with near-degenerate spin states.~\cite{Dobrautz2019, dobrautz-phd}
However, compared to the SD-based FCIQMC, the calculation of (spin free) RDMs in GUGA-FCIQMC is considerably more challenging, and hence such RDMs have not been available until now. This has prevented to access properties,
and using it as a spin-pure CI-solver within the Stochastic-CASSCF method.~\cite{LiManni2016}

\section{\label{sec:guga-rdms}GUGA-RDMs}
In this section the theoretical and algorithmic details of 
the stochastic sampling of RDMs within the GUGA-FCIQMC method 
are discussed. 

\subsection{Theoretical considerations}
Unbiased RDM sampling within the FCIQMC algorithm,  whether in SD 
or CSF basis, is made possible by  the \emph{replica} method, 
where two independent dynamics are simultaneously carried to 
remove a strictly positive bias due to stochastic fluctuations 
for the diagonal RDM contributions.~\cite{Alavi2014}

In a SD based implementation the 1-particle RDM entries
\begin{equation}
\begin{aligned}
\rho_{ij,\sigma}(\tau) &= \braopket{\Psi(\tau)}{\hat a_{i\sigma}^\dagger \hat a_{j\sigma}}{\Psi(\tau)} \\ &=
\sum_{IJ} c_I^\text{A}(\tau) c_J^\text{B}(\tau) \braopket{I}{\hat  a_{i\sigma}^\dagger \hat a_{j\sigma}}{J},
\end{aligned}
\end{equation}
are derived from the stochastic coefficients $c_I^\text A(\tau)$ and $c_J^\text{B}(\tau)$ 
of two statistically independent calculations, $A$ and $B$.
The two-body RDMs are obtained in a similar way.
For SDs, the terms $\braopket{I}{a_{i\sigma}^\dagger a_{j\sigma}}{J}$ are
promptly given by the well-known Slater-Condon rules.
We make use of the fact that $\ket{I} \ra \ket{J}$ transitions are
already performed in FCIQMC during the stochastic \emph{spawning step}.
Hence, we reuse the information, already required for a \emph{normal} simulation, 
to additionally sample the 1- and 2-RDM elements.

In the original SD based implementation this is done by additionally storing
information of the \emph{parent} SD, $\ket{J}$, along with the \emph{spawned} new 
SD, $\ket{I}$, including the parent SD encoded in a bit representation, 
its coefficient, in what run (A or B) this spawn happened,
and other implementation specific flags.

In a parallel high-performance computing (HPC) environment, 
the occupied determinants are distributed among the different processors.
Hence, the newly spawned walkers  are kept in an array, which
has to be communicated to the corresponding processor, 
where the newly spawned state is stored, to update the corresponding coefficients.

The spin-free one- and two-body
RDMs in terms of unitary group generators~\cite{Luzanov1985} 
are defined as (following the convention of Helgaker, J{\o}rgensen and Olsen~\cite{Helgaker2000})
\begin{equation}
\label{eq:one-rdm}
\rho_{ij} = \braopket{\Psi}{\hat E_{ij}}{\Psi} = \sum_{\mu\nu} c_\mu^* c_\nu \braopket{\mu}{\hat E_{ij}}{\nu},
\end{equation}
with $\hat E_{ij}^\dagger = \hat E_{ji}$, and
\begin{equation}
\label{eq:two-rdm}
\begin{aligned}
\Gamma_{ij, kl} &= \braopket{\Psi}{\hat e_{ij, kl}}{\Psi} = \sum_{\mu\nu} c_\mu^* c_\nu \braopket{\mu}{\hat e_{ij, kl}}{\nu} \\ &= \sum_{\mu\nu}c_\mu^* c_\nu \braopket{\mu}{\hat E_{ij} \hat E_{kl} - \delta_{jk}\hat E_{il}}{\nu},
\end{aligned}
\end{equation}
with $i, j, k, l$ denoting \emph{spatial} orbitals, $\ket{\mu}$ and $\ket{\nu}$ being configuration state 
functions (CSFs) and $c_\mu$ and $c_\nu$ their coefficients in the ground state wave function expansion, $\ket{\Psi}$.

The diagonal terms of the RDMs are accumulated \emph{explicitly}, and the diagonal 1-RDM terms reduce to
\begin{equation}
\label{eq:diag-1-rdm}
\rho_{ii} = \sum_{\mu} c_\mu^{\text A} c_\mu^{\text{B}} \braopket{\mu}{\hat E_{ii}}{\mu} = \sum_\mu c_\mu^{\text A} c_\mu^{\text{B}}\, n_i,
\end{equation}
where $n_i$ is the occupation of the \emph{spatial} orbital $i$, 
which can assume the values $0$, $1$, or $2$.
The diagonal 2-RDM elements are defined as
\begin{align}
\label{eq:diag-2-rdm}
\Gamma_{ii, jj} &= \sum_\mu  c_\mu^{\text A} c_\mu^{\text{B}} \, \braopket{\mu}{\hat e_{ii, jj}}{\mu} \nonumber\\
& = \sum_\mu c_\mu^{\text A} c_\mu^{\text{B}} \,
\braopket{\mu}{\hat E_{ii}\hat E_{jj} - \delta_{ij}\hat{E}_{ij}}{\mu},
\end{align}
which for $i = j$ yields
\begin{equation}
\label{eq:diag-2-rdm-ii}
\Gamma_{ii, ii} = \sum_\mu  c_\mu^{\text A} c_\mu^{\text{B}} 
\,n_i (n_i - 1)
\end{equation}
and for $i \neq j$
\begin{equation}
\label{eq:diag-2-rdm-ij}
\Gamma_{ii, jj} =  \sum_\mu  c_\mu^{\text A} c_\mu^{\text{B}}  n_i \, n_j.
\end{equation}
Eqs.\eqnref{eq:diag-2-rdm-ii} and\eqnref{eq:diag-2-rdm-ij} are simply 
products of orbital occupation numbers and the coefficients $c_\mu^{\text{A/B}}$ from two
statistically independent simulations, due to the above mentioned positive bias in diagonal RDM entries.
Exchange type elements of the 2-RDM, $\Gamma_{ij,ji}$, also have diagonal contributions from the wave function 
\begin{equation}
\Gamma_{ij,ji} = \sum_\mu c_\mu^A c_\mu^B \braopket{\mu}{\hat e_{ij,ji}}{\mu},
\end{equation}
which are also sampled explicitly. The detailed form of the coupling coefficients can be found in the literature.~\cite{Shavitt1978, Dobrautz2019} 
These exchange-like terms do, however, also have off-diagonal contributions, $\braopket{\mu}{\hat e_{ij,ji}}{\nu}$, which are explained in general in the following section.

\subsection{Off-diagonal RDM entries -- computational implementation and cost}

Similar to the SD based RDM sampling, for each sampled RDM element we store 
the parent state, $\ket{\mu}$, its coefficient, $c_\mu$, and the replica index ($A$ or $B$).
However, there are some important differences in the GUGA based RDM sampling
compared to a SD-based implementation:

\textbf{(a)} The one-electron \emph{coupling coefficients}
$\braopket{\mu'}{\hat E_{ij}}{\mu}$, and the corresponding two-body terms, 
$\braopket{\mu}{\hat e_{ij,ji}}{\nu}$, 
do not follow the Slater-Condon rules as for SDs.
Shavitt and Paldus derived an efficient \emph{product} form 
of these coupling coefficients, exemplified by a single excitation as,
\begin{equation}
\label{eq:prod-form}
\braopket{\mu'}{\hat E_{ij}}{\mu} = \prod_{k=i}^j W(d_k', d_k, S_k),
\end{equation}
where $W$ is a function of the \emph{step-values}, $d_k = \{0,u,d,2\}$, 
of spatial orbital $k$ of the step-vector representation of the two CSFs, 
$\ket{\mu}$ and $\ket{\mu'}$, and the intermediate value of the total spin, 
$S_k$, in the cumulative sense.
The step-values, $d_k$, encode if a spatial orbital is empty, $d_k = 0$, 
positively spin-coupled $\Delta S_k = +1/2$, $d_k = u$, 
negatively spin-coupled, $d_k = d$ or doubly occupied, $d_k = 2$.
CSFs can be represented graphically (see Fig.~\ref{fig:exchange-example}a)
where different step-values are indicated by a different tilt
of the segments, and Shavitt showed that the value of the coupling coefficients 
only depends on the \emph{loop shape} enclosed by the two coupled CSFs.

Their explicit calculation scales with the number of
spatial orbital indices between $i$ and $j$. 
However, we calculate this quantity on-the-fly, during the excitation 
generation step, and thus, we can reuse it with
no additional computational 
cost in the stochastic RDM sampling.

\textbf{(b)} \emph{Identifying} the type of excitation
and the involved \emph{spatial} orbitals $(i, j, k, l)$, when 
coupling CSFs, is a more complex operation than for SDs.
CSFs can also differ in the open-shell spin-coupling,
and not only in the specific spatial orbitals $(i, j, k, l)$, 
yet still have a non-zero coupling coefficient.
For example, in Figure~\ref{fig:exchange-example}a we show  
Shavitt's graphical representation of the CSF $\ket \mu = \ket{u,0,u,d,u,0}$  (already used in Fig.~\ref{fig:branching-tree}) as the orange solid line and an excited CSF $\ket \mu' = \ket{u,0,2,u,0,,0}$ as the blue dashed line. 
Following Shavitt's convention the CSFs are drawn from bottom to top.
Only the gray \emph{loop} area enclosed by both CSFs contributes to the coupling coefficient, $\braopket{\mu}{E_{ij}}{\mu'}$. 
The two CSFs are connected by an excitation of an electron from orbital $j=5$ to $i = 3$, indicated by the arrow. However, as one can see in Fig.~\ref{fig:exchange-example}a, the two CSFs $\ket{\mu}$ and $\ket{\mu'}$ do also differ in the spin coupling of orbital $k = 4$ with $d_k = d$, while $d'_k = u$ (in the step-value notation).
Hence, it is not as simple as performing bit-wise logical operations on 
alpha- and beta-strings as it is possible for SDs,~\cite{Giner2013}
to identify the involved spatial indices and type of excitation.
We do have optimized routines to perform this excitation identification 
for arbitrary CSFs in our GUGA-FCIQMC code~\neci,~\cite{Guther2020,neci}
and similar to the above-mentioned coupling coefficients,
we already have the necessary information in the excitation 
process, within the spawning step.

\textbf{(c)} Certain excitation types, such as the exchange-like excitations,
$\hat e_{ij,ji}$ and $\hat e_{ij,jk}$, can have multiple non-unique spatial 
orbital combinations leading to the same type of excitation $\ket \mu \ra \ket{\mu'}$.
This stems from the fact that certain contributions to the two-body coupling 
coefficients, $\braopket{\mu}{\hat e_{ij,kl}}{\mu'}$, are non-zero for alike 
open-shell step-values, $d_o = d_o'$, above and below the loop spawned 
between $\ket{\mu}$ and $\ket{\mu'}$, see Figure~\ref{fig:exchange-example}b and Shavitt.~\cite{Shavitt1978}

For example, for a pure exchange-type excitation, $\hat e_{ij,ji}$, as 
depicted in Figure~\ref{fig:exchange-example}b,
only the spin-coupling of the open-shell orbitals differs, 
but there is no change in the orbital occupation.
To calculate the Hamiltonian matrix element, $\braopket{\mu}{\hat H}{\mu'} =
\sum_{i\neq j} V_{ijji} \braopket{\mu}{\hat e_{ij,ji}}{\mu'}$,
one needs to consider all non-zero contribution to the coupling coefficient, 
from orbital $i'$ below and $j'$ above the loop.
Additionally, as the specific spatial orbitals, $i,j$ ($k,l$) are 
chosen \emph{first} in the excitation generation in FCIQMC, it is 
necessary to also take into account the possibilities that the
other contributing orbitals, $p(i',j')$, would have been picked 
(as their choice could have led to the same excitations), 
to assign a \emph{unique} total generation probability, $p_{gen}(\mu'\vert \mu)$.
However, for a correct RDM sampling we have to retain the \emph{original} probability
$p(\mu \ra \mu' \vert i, j, k, l)$ to sample a specific $\Gamma_{ij, kl}$ 
entry to avoid a possible double counting.
Conveniently, similar to the cases \textbf{(a)} and \textbf{(b)} mentioned above, 
we already have access to this specific quantity, obtained during the excitation 
generation process and do not need to explicitly recalculate it for the 
stochastic RDM sampling.

The \emph{three} additional necessary quantities discussed above, namely 
the coupling coefficient, $\braopket{\mu'}{\hat E_{ij}}{\mu}$ or $\braopket{\mu'}{\hat e_{ij,kl}}{\mu}$,
the excitation type, and 
the probability $p(\mu\ra \mu'\vert i, j, (k, l))$ 
are already computed in the random excitation process.
Consequently, the main change to enable spin-free RDM sampling within GUGA-FCIQMC is 
to communicate these three additional quantities, along with the already 
communicated information of the parent state, $\ket{\mu}$, its coefficient, $c_\mu$, 
and the replica index, $A/B$.

An important algorithmic advancement and routinely used feature of FCIQMC was the \emph{semi-stochastic} method,~\cite{Blunt2015-semistoch, Umrigar2012} where some chosen part of the Hilbert space
-- usually the $N_D$ most occupied states -- is treated explicitly.
This is achieved 
by constructing the full Hamiltonian matrix $H_{\mu\nu}, \forall \mu,\nu \in \{N_D\}$ and 
performing the imaginary-time evolution exactly.
This necessitates also a change to the RDM sampling, since the RDMs contributions from states within the semi-stochastic space are not covered in the random excitation process anymore.
These RDM contributions are treated exactly, greatly increasing their accuracy on the one hand,
but -- especially in the spin-free case -- also increasing the computational effort.
In this case it is not possible to avoid the explicit \emph{excitation identification}
and \emph{coupling coefficient} and \emph{original generation probability calculation} in GUGA-FCIQMC.
However,
there is only marginal computational overhead of around 10-20\% associated with the spin-free RDM sampling compared to a standard two-replica FCIQMC calculation (see the SI for details).

The spin-adapted Stochastic-CASSCF method has 
been made available in the \texttt{OpenMolcas} chemistry software package.~\cite{OpenMolcas2019}

\section{\label{sec:results}Results and Discussion}

\begin{figure}
	\includegraphics[width=0.4\textwidth]{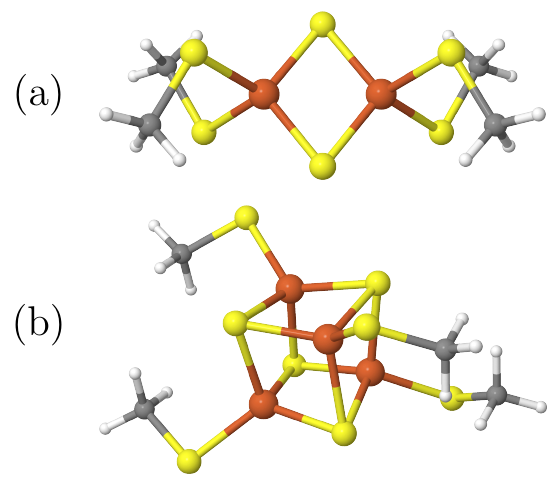}
	\caption{\label{fig:fe2s2-model}Geometry of the (a) [Fe$_2$S$_2$(SCH$_3$)$_4]^{2-}$ model system derived from synthetic complexes of Mayerle \emph{et al.}~\cite{Mayerle1973, Mayerle1975} and (b) [Fe$_4$S$_4$(SCH$_3)_4)]^{2-}$ model system
		obtained from synthetic complexes of Averill \emph{et al.}.~\cite{Averill1973}
		Orange indicates iron, yellow sulfur, gray carbon and white hydrogen atoms.}
\end{figure}

The GUGA-FCIQMC RDM sampling has been used within the Stochastic-CASSCF framework
to study the low-energy spin states of the $\ce{[Fe(III)_2S_2(SCH3)4]^{2-}}$ model
complex (Figure~\ref{fig:fe2s2-model}a), derived from synthetic complexes of 
Mayerle \emph{et al.},~\cite{Mayerle1973,Mayerle1975} and utilized in our previous
investigation,~\cite{LiManni2020}
and the $\ce{[Fe(III)_4S_4(SCH3)_44]}$ model cubane (Figure~\ref{fig:fe2s2-model}b),
obtained from the synthetic complex of Averill~\emph{et al.},~\cite{Averill1973}
where the terminal groups have been replaced by methyl groups.
For the $\ce{[Fe2S2]}$ system, we considered 
\textbf{(1)} a CAS(10e,10o), consisting of the singly occupied iron $\ce{3d}$ orbitals, 
\textbf{(2)} a CAS(10e,20o), consisting of the singly occupied iron $\ce{3d}$ and the
empty correlating \textit{double-shell} $\ce{d}'$ orbitals,
\textbf{(3)} a CAS(22e,16o) consisting of the singly occupied iron $\ce{3d}$ and the six doubly 
occupied bridging sulfur 3p orbitals,
and \textbf{ (4)} a CAS(22e,26o) containing the iron $\ce{3d}$ and $\ce{d}'$ orbitals,
and the six bridging sulfur $\ce{3p}$ orbitals. This
active space correspond to the one
utilized in our previous work.~\cite{LiManni2020}

We also studied the role of the iron $\ce{4s}$ and the peripheral sulfur $\ce{3p}$ orbitals,
which were considered in other studies of similar $\ce{FeS}$ 
dimers,~\cite{Sharma2014,Cho2019,Chilkuri2019}
having mixed-valence states as main target.
We found that the iron $\ce{4s}$ orbitals have a negligible differential role on the low-energy spin gaps.

Including one terminal orbital per peripheral sulfur atom in the active space resulted
in an uneven mixing between different orbitals on some of the peripheral sulfur
atoms upon completion of the CASSCF procedure.
This suggests that for a balanced treatment of the peripheral S orbitals 
one would need to include all 12 of them. However, while these orbitals have 
important ligand field effects that could affect the energetic of mixed-valence
states, we found that their role is less crucial for dealing with the homo-valent 
$\ce{[Fe(III)S]}$ systems.
Additionally, a recent study on the excited state spectrum of the $\ce{[FeS]}$ dimer~\cite{Kubas2020} using the 
CAS(22e,16o) wave functions,
showed that the low-lying non-Hund excited states involve
bridging-sulfur charge transfer (CT) states, while CT states involving terminal-sulfur orbitals were only found at higher energies. 

Thus, we decided not to further consider these orbitals in the chosen model
active space. This was considered to be a successful strategy in 
previous works.~\cite{Presti2019, LiManni2020b, LiManni2020}
Similar to previous computational studies~\cite{LiManni2020b,LiManni2020,Sharma2014,Cho2019,Chilkuri2019,Presti2019} 
we do not include \emph{empty} sulfur orbitals in our active space.
Therefore, metal-to-ligand charge transfer (MLCT) excitations are not considered 
by the model active space chosen. However, as also suggested by Neese
\emph{et al.},~\cite{Chilkuri2019} such configurations are rather high in energy,
and they can be safely neglected for low-energy spectrum calculations.

We used an extended relativistic atomic natural orbital basis of
double-$\zeta$ quality for Fe atoms and a minimal basis 
for all other elements. 
The exactly diagonalizable Fe$_2$S$_2$ (10e,10o), (10e,20o) and (22e,16o) active 
spaces are straightforwardly calculable within spin-adapted Stochastic-CASSCF with 
modest computational resources. 
We ensured the convergence of the (22e,26o) active space calculations with respect to the 
number of walkers, $N_w$, by increasing $N_w$ up to $N_w = 1\cdot10^9$, see the SI\cite{SI} for more information.
The average number of occupied CSFs, $N_{CSF}$, at each time-step during the GUGA-FCIQMC calculation 
for $N_w = 5\cdot 10^8$ and each spin state is shown in Table~\ref{tab:occ-csfs}.
The size of the deterministic space, $N_D$, which is treated exactly within GUGA-FCIQMC was $N_D = 5\cdot 10^4$ for these 
calculations.
With the spin-adapted implementation of FCIQMC via the GUGA, 
wavefunctions containing hundred of millions of CSFs (with many open-shell orbitals) can be efficiently treated. 
Detailed further information on the geometries, orbitals and computations
can be found in the SI.

\begin{table}
	\caption{\label{tab:occ-csfs}
		Average number of occupied CSFs, $N_{CSF}$ (in millions), for each spin state in the Fe$_2$S$_2$ (22e,26o) active space calculations with $N_w = 5\cdot 10^8$.
	}
	{\small 
		\begin{tabular}{ccccccc}
			\toprule
			Total spin				&  0	&  1	&  2  &  3 & 4 & 5 \\
			\midrule
			$N_{CSF}$ in millions	&	333 	& 345   & 338 & 324 & 300 & 268 \\
			\bottomrule
		\end{tabular}
	}
\end{table}

\subsection{Fe$_2$S$_2$ system \label{sec:fe2}}

The (10e,10o), (10e,20o) and (22e,16o) active spaces are exactly diagonalizable,
and were considered to study the differential interplay of different correlation mechanisms,
such as orbital relaxation, \textit{double-shell}~\cite{LiManni2018, Veryazov2011, Fulde2014, Andersson1992} 
and superexchange~\cite{Kramers1934, Anderson1950, Goodenough1955, Kanamori1959,Bogdanov2018} correlation effects,
and to benchmark and test our stochastic spin-free RDM sampling procedure.
A thorough comparison of the exact and the Stochastic-CASSCF results can be found in the SI~\cite{SI}.

In our earlier works,~\cite{LiManni2020,LiManni2020b} we have 
demonstrated via theoretical arguments, and shown with calculations, 
that  the choice of orbital representation and reordering greatly effect the 
sparsity of the CI wave function within the GUGA formalism.
We have also shown that the localization and reordering strategy within the 
GUGA-FCIQMC algorithm is of utmost importance, as it positively influences the stability of 
the dynamics and the convergence with respect to the total number of walkers.
Moreover, this strategy greatly simplifies the interpretation of the converged wave 
functions, and could even allow selective optimization of one among ground- and 
low-energy excited-state wave functions.
We have adopted the same strategy for the present work.
In Reference~[\citen{LiManni2020}] the optimized CASSCF(22e,26o) orbitals for the $S=0$ ground state, 
obtained via the SD-based Stochastic-CASSCF,~\cite{LiManni2016} were used as starting orbitals
for the localization and reordering protocol and for the GUGA-FCIQMC dynamics.
A CASSCF(10e,10o) was performed inside the CAS(22e,26o) active space, an invariant rotation 
within the CAS(22e,26o), that separates valence $\ce{3d}$ orbitals from the six sulfur 
and the 10 correlating $\ce{d}'$ orbitals.
Only the 10 valence $\ce{3d}$ orbitals were localized and site-ordered, 
leaving the sulfur and the correlating $\ce{d}'$ orbitals delocalized.
In the present work, the starting orbitals were obtained from a high-spin
restricted open-shell Hartree-Fock (ROHF) calculation, equivalent to a 
CASSCF(10e,10o) $S = 5$ optimization. 
The iron $\ce{3d}$ and $\ce{d}'$ orbitals, resulting from the ROHF calculation, were
separately localized, using the Pipek-Mezey~\cite{Pipek1989} method,
while the bridging sulfur $\ce{3p}$ orbitals were left delocalized.
Using localized $\ce{d}'$ orbitals allows to better estimate the local spin of each magnetic center.

\begin{figure}
	\includegraphics[width=0.5\textwidth]{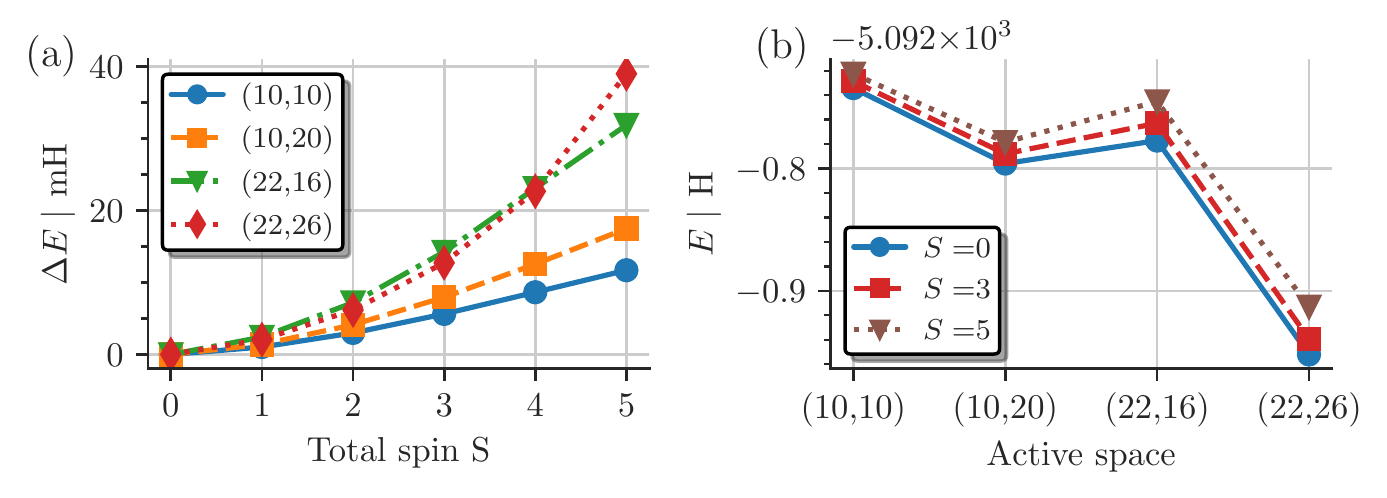}
	\caption{\label{fig:spin-gaps-and-tot-e-final}(a) CASSCF spin gaps relative to the $S = 0$ state as a function of total spin $S$ for different active spaces and (b) total CASSCF energies of the $S = 0, 3$ and $5$ states as a function of the active spaces. 
	}
\end{figure}

\textbf{Fe$_2$S$_2$ spin ladder and total energies}\\
Figure~\ref{fig:spin-gaps-and-tot-e-final}a
shows the spin gaps of all the states relative to the $S = 0$ ground state -- the \emph{spin ladder} --
as function of the total spin after the CASSCF orbital optimization.
The spin gaps are lowest in the (10e,10o) active space, with $\Delta E = 12$~mH, between the $S = 5$ and $S = 0$ state. 
The inclusion of the iron $\ce{d}'$ orbital in the (10e,20o) active space qualitatively does not change the 
obtained spin ladder and it also has a rather smaller quantitative effect, with an  only slightly larger 
$\Delta E = 17$~mH between the $S = 5$ and $S = 0$ state. 
Inclusion of the bridging-sulfur $\ce{3p}$ orbitals has the largest effect on the spin gaps,
since it accounts for the metal-bridging ligand correlation, which is differentially more important
than the radial correlation effect,~\cite{Fulde2014} accounted for
by the inclusion of the $\ce{d}'$.
The consideration of both the iron $\ce{d}'$ and the bridging-sulfur $\ce{3p}$  orbitals in the (22e,26o) 
active space induces a qualitative change in the obtained spin gaps, which will be further discussed below.
Quantitatively the relative spin gaps enlarge by as much as a factor of 3.3, 
when enlarging the active space, from CAS(10e,10o) to CAS(22e,26o).

In Figure~\ref{fig:spin-gaps-and-tot-e-final}b  we show the total energy of the $S = 0$, $3$ and $5$ states,
helpful in describing in absolute terms the correlation effects bound to ligand-to-metal charge-transfer 
and radial correlation effects. 
Starting from the CAS(10e,10o), the inclusion of the iron correlating $\ce{d}'$ orbitals, 
as in the CAS(10e,20o) active space, lowers the total energy more than including the 
sulfur $\ce{3p}$ orbitals, as in the CAS(22e,16o).
The combined inclusion of both iron $\ce{d}'$ and sulfur $\ce{3p}$ orbitals 
has the surprising effect of lowering the total energies more than the 
ligand-to-metal charge-transfer and the radial correlation effects on their own.
However, the largest differential effect arises from the ligand-to-metal charge-transfer
excitations as show in Figure~\ref{fig:spin-gaps-and-tot-e-final}a.

\textbf{Orbital relaxation effect.} \\
In this section the overall and the differential effect 
of the CASSCF orbital relaxation on energies and spin-gaps, together with
its effect on the derived model parameters, is discussed.
The highest-spin, $S=5$, restricted open-shell Hartree-Fock (ROHF) orbitals from the (10e,10o)
active space
are chosen as starting orbitals for all the calculations. 
The results of the first CASSCF iteration are from here on referred to as \emph{CASCI}.

\begin{figure}
	\includegraphics[width=0.5\textwidth]{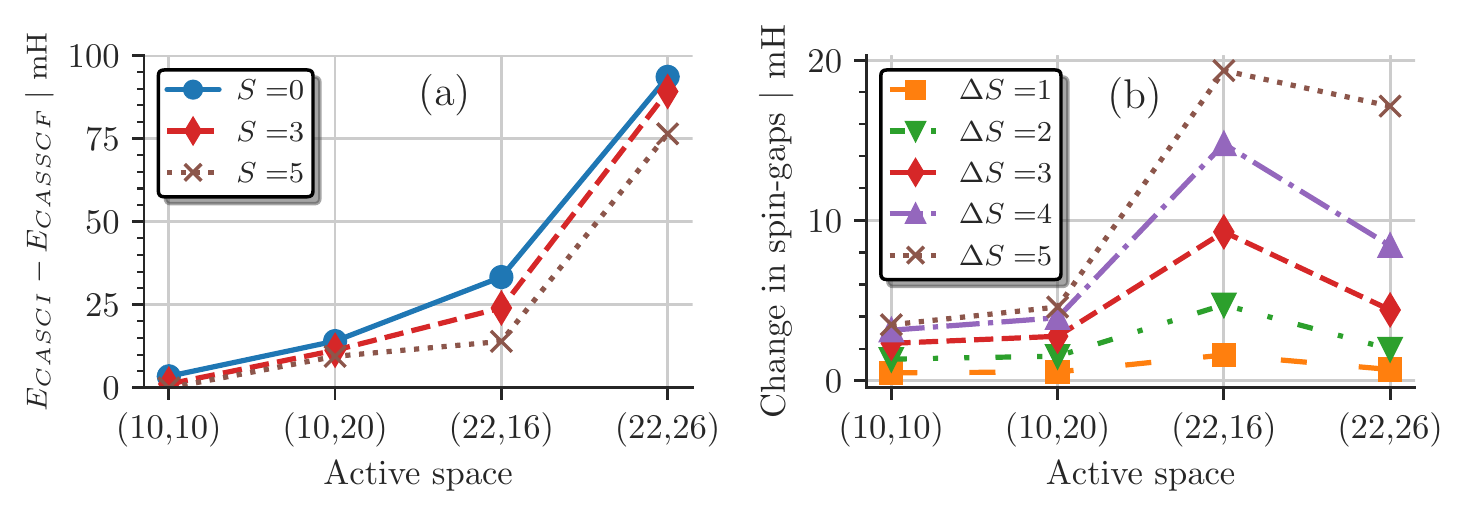}
	\caption{\label{fig:tot-e-orb-relax}(a) Change of the total energy for the $S=0,3$ and $5$ states 
		and (b) change of the spin gaps relative to the $S=0$ state due to the CASSCF orbital relaxation
		as a function of active space using (10e,10o) ROHF as starting orbitals (CASCI).
	}
\end{figure}

Figure~\ref{fig:tot-e-orb-relax}a shows 
the energy difference of the CASCI results using (10e,10o) ROHF
orbitals and the CASSCF results, $\Delta E = E_{\text{CASCI}} - E_{\text{CASSCF}}$,
for the $S = 0, 3$ and $5$ states as a function of the active space. 
As expected, the effect of the CASSCF orbital relaxation, when using 
(10e,10o) ROHF orbital, is lowest for the (10e,10o) active (with differences below 10 mH), 
and highest for the (22e,26o) active space. 
Within each active space the effect of the CASSCF procedure is largest for the low spin 
states, with a maximum difference of $\Delta E = 94$~mH for the singlet in the (22e,26o)
active space. 
For the high spin states the effect of the CASSCF procedure is smaller, but 
still substantial for the larger active spaces, especially in the (22e,26o) AS, with $\Delta E = 76$~mH for the $S = 5$ state.

To investigate the differential effect, Figure~\ref{fig:tot-e-orb-relax}b shows the changes in the spin gaps,
due to CASSCF orbital relaxation as a function of active space.
As expected, the CASSCF procedure increases all the obtained spin-gaps, as
the low spin states are stabilized more by the orbital relaxation
when starting from high-spin ROHF orbitals than the higher spin-states,
which are better represented by the ROHF orbitals.
The change in the spin-gaps is smallest for the (10e,10o), 
where the ROHF starting orbitals were obtained, 
and the somehow similar (10e,20o) active space.
Interestingly, 
although the effect of the CASSCF procedure on the total energies is highest for the 
(22e,26o) active space, see Fig.~\ref{fig:tot-e-orb-relax}a, the largest effect on the spin gaps 
is observed in the intermediate (22e,16o) active space.
The energy differences to low-spin states, $\Delta S = {1, 2, 3}$, are affected only weakly
by the CASSCF optimization and stay similar to the CASCI results.
This can be explained by the fact that the low-spin states are 
similarly biased in the ROHF orbital basis, and thus show 
similar stabilization during the CASSCF procedure.
The high-spin states, on the other hand, are less stabilized by the CASSCF procedure,
and as a consequence the low-to-high spin-gaps result are enlarged 
by the orbital relaxation.

There is a substantial differential effect of $\approx 15-20$ mH 
due to the CASSCF orbital relaxation, 
so one has to be cautious when using ROHF orbitals for spin systems,
and orbital bias towards the high-spin is to be expected, leading to
a systematic under-estimation of spin-gap predictions for 
anti-ferromagnetically coupled magnetic sites.
Even for the seemingly SCF-invariant singlet-triplet spin gap,
the CASSCF procedure is crucial to obtain
more accurate model magnetic parameters, as it will be discussed below.

\begin{figure}
	\includegraphics[width=0.35\textwidth]{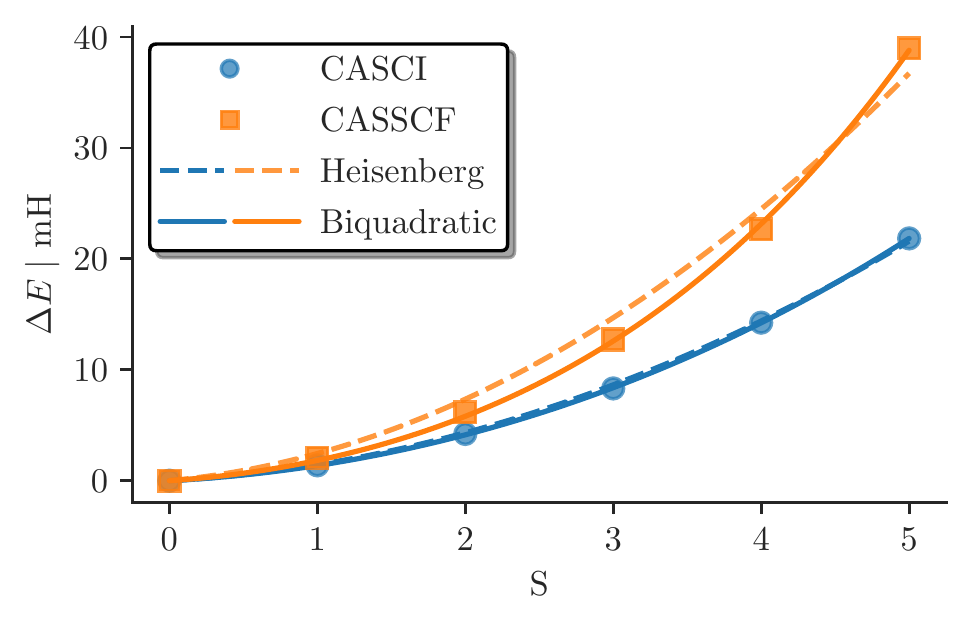}
	\caption{\label{fig:fe2-22in26-spinladder}
		Energy difference to the $S = 0$ ground state as a function of spin in the (22e,26o)
		active space for the \emph{ab initio} CASCI (blue) and CASSCF (orange)
		results with a simple (dashed line) and biquadratic (solid line) Heisenberg model fitted to the data.}
\end{figure}

Figure~\ref{fig:fe2-22in26-spinladder} shows the energy differences
with respect to the $S=0$ ground state, 
for the CASCI(22e,26o) (blue circles), and for the CASSCF(22e,26o) (orange square) results.
As expected, the spin states are more separated after the CASSCF 
orbital optimization, with a lowest-to-highest spin-state gap 
nearly doubled by the orbital relaxation effects.
Figure~\ref{fig:fe2-22in26-spinladder}  also shows the spin ladder obtained from mapping
the \emph{ab initio} results to a Heisenberg
model,~\cite{Heisenberg1928,Dirac1926,Kittel1960,Anderson1959,Boca1999}
with (dashed lines) and without (solid lines) a \textit{biquadratic}
correction. This aspect will be discussed in greater detail in the 
next section.

\textbf{Mapping to a spin model}\\
As previously done by
Sharma \etal~\cite{Sharma2014} and in our laboratories,~\cite{LiManni2020b}
we map the ab initio low-energy spectrum of the $\ce{Fe2S2}$ system to a spin 
Hamiltonian, as the spin-exchange interactions are the dominant
form of magnetic interactions in this system.
First we map the excitation energies of the $\ce{Fe2S2}$ system to 
the \emph{bilinear} two-site Heisenberg Hamiltonian
\begin{equation}\label{eq:simple-heisenberg}
\hat H = J \, \hat{\mathbf{S}}_A \cdot \hat{\mathbf{S}}_B, 
\end{equation}
with eigenvalues
\begin{equation}\label{eq:HeisenbergEigenVal}
\quad E(S) =  \frac{J}{2} S (S + 1),
\end{equation}
where $\hat{\mathbf{S}}_{A/B}$ are the local spin-$\nicefrac{5}{2}$
operators of the two iron centers, and $S$ is the total targeted spin.
We obtain the magnetic coupling  parameter $J$ by performing a least-squares 
fit of the energy expression, Equation~(\ref{eq:HeisenbergEigenVal}), 
to the \emph{ab initio} results of all lowest spin states and 
study the quality of this mapping as a function of the active space 
size and the effect of the CASSCF orbital optimization.
As shown in Figure~\ref{fig:fe2-22in26-spinladder}, the bilinear Heisenberg spin-ladder (solid blue line) 
models  the \emph{ab initio} 
CASSCI results with high accuracy.
However, minor deviations can be observed for the fitting of the CASSCF results. 
This finding suggests that orbital relaxation effects account for additional forms of 
interactions between the metal centers in addition to enlarging the predicted $J$ values.
An improved Heisenberg model with \emph{biquadratic} 
exchange,~\cite{Boca1999,Malrieu2008,Kittel1960,Anderson1959,Falk1984,
	Bastardis2008,Calzado1998,Bastardis2007,Moreira2002}
\begin{equation}\label{eq:biquad-1}
\hat H = J' \,\hat{\mathbf{S}}_A \cdot \hat{\mathbf{S}}_B + K \left( \hat{\mathbf{S}}_A \cdot \hat{\mathbf{S}}_B \right)^2,
\end{equation}
with eigenvalues
\begin{equation}\label{eq:biquad-energy}
\begin{aligned}
E(S) = \frac{J'}{2} S ( S +& 1) + \frac{K}{4}  S ( S + 1) \Bigl[S (S + 1) + 1  \\ & - 2 S_A (S_A + 1) - 2 S_B (S_B + 1)\Bigr],
\end{aligned}
\end{equation}
greatly improves the fitting of the model Hamiltonian (dashed lines in Figure~\ref{fig:fe2-22in26-spinladder}).

\begin{figure}
	\includegraphics[width=0.5\textwidth]{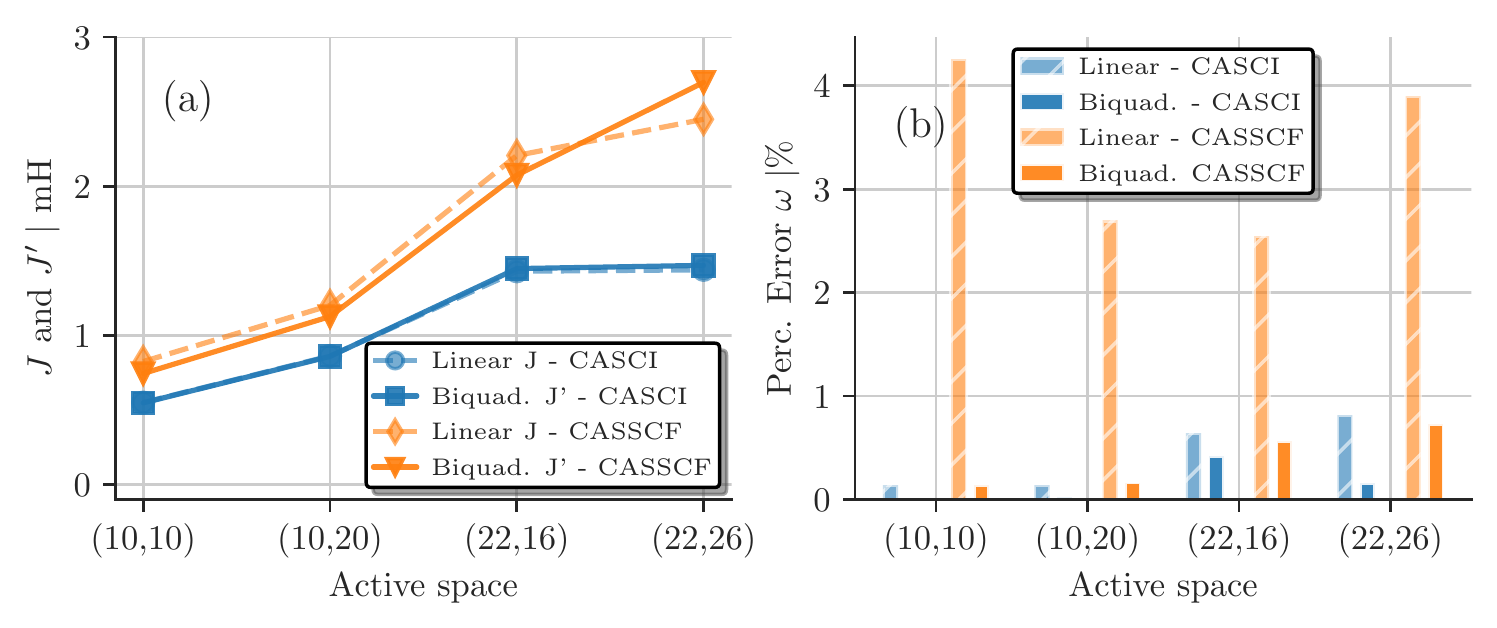}
	\caption{\label{fig:heisenberg-vs-cas}(a) Bilinear Heisenberg $J$ (dashed lines) and biquadratic $J'$ (solid lines) fit of the \emph{ab initio} CASCI (blue) and CASSCF (orange) results as a function of the 
		active space sizes. 
		(b) relative average error per sate $\omega$ in percent of the corresponding bilinear (dashed) and biquadratic (solid) Heisenberg fits of (a) as a function of the active space.
	}
\end{figure}

Figure~\ref{fig:heisenberg-vs-cas}a shows the fitted model parameters of the 
bilinear, $J$ (Eq.~\eqref{eq:simple-heisenberg}), and biquadratic, $J'$ (Eq.~\eqref{eq:biquad-1}),
Heisenberg model as a function of the active space size for the CASCI (blue squares and circles) and CASSCF (orange triangles and diamonds) results.
For the CASCI results (blue), the extracted model parameters, $J$ and $J'$, are almost identical for all active 
spaces, indicating a good description by the bilinear Heisenberg model.
$J$ and $J'$ increase from a value of $0.55$~mH in the (10e,10o) active space to about $1.44$~mH in the (22e,16o) and 
(22e,26o) AS. 

For the CASSCF results (orange),
the extracted $J$ (diamonds) and $J'$ (triangles) parameters are larger than the corresponding CASCI results, increasingly so in the larger active spaces, and additionally, the bilinear $J$ and biquadratic $J'$ differ.
For all but the largest active space, the biquadratic $J'$ is about 0.1~mH smaller than the bilinear $J$, while 
it is $\sim0.25$~mH larger in the (22e,26o) AS.
The differences between the extracted model parameters indicate that a simple bilinear Heisenberg model is not 
sufficient to describe the  CASSCF results.

To quantify this discrepancy and analyze how well a biquadratic model suits the \emph{ab initio} results we    
show the relative average error per state $\omega$ (in percent) of the 
corresponding bilinear and biquadratic Heisenberg fits to the CASCI (blue solid and striped) and CASSCF (orange solid and striped) results in Figure~\ref{fig:heisenberg-vs-cas}b .
Following Ref.~[\citen{Malrieu2008}], $\omega$ is defined as
\begin{equation}\label{eq:perc-error}
\omega = \frac{100}{N \Delta E_{max}^C} \sum_{S=1}^N \abs{E_S^C - E_S^M},
\end{equation}
where $E_S^C$ is the computed \emph{ab initio} spin gap of spin state $S$ relative to the singlet ground state and $E_S^M$ is 
the energy obtained by fitting the bilinear and biquadratic model, Eqs.\eqnref{eq:HeisenbergEigenVal} and~\eqref{eq:biquad-energy}.
$N$ is the number of considered states (with $N = 5$ in the FeS dimer case, as we only consider the spin-gap relative to the singlet ground state) and $\Delta E_{max}^C$ is the \emph{ab initio} energy difference between the $S = 5$ and singlet state.

The CASCI spin ladders exhibit a clear bilinear Heisenberg behavior,
as shown by the small $\omega$ values (blue bars) in Fig.~\ref{fig:fe2-22in26-spinladder}b.
The error is less than $1\%$ for all active space sizes.
Larger discrepancies emerge between the CASSCF energies and 
the bilinear Heisenberg model, indicated by larger $\omega$ values (orange striped bars in Fig.~\ref{fig:fe2-22in26-spinladder}b).
The relative error $\omega$, defined in Eq.~\eqref{eq:perc-error} takes into account the 
gap between the $S = 0$ and $S = 5$ state, $\Delta E_{max}^C$, in the denominator.
This causes $\omega$ to be largest for the bilinear Heisenberg fit to the CASSCF results in the (10e,10o) active space.

Overall, these discrepancies are still rather small (at most $4\%$) as shown in Figure~\ref{fig:fe2-22in26-spinladder}b, however, not negligible.
The biquadratic Heisenberg Hamiltonian, Eq.\eqnref{eq:biquad-1}, 
describes the CASSCF results better, as indicated by 
a much smaller $\omega$ value (less than 1\%) in all cases, see Fig.~\ref{fig:fe2-22in26-spinladder}b.
However, the largest CAS(22e,26o) starts to show deviations also from 
the biquadratic Heisenberg model.
Independently of the quantitative aspects, our calculations confirm the anti-ferromagnetic character
of this system, with the CASSCF predicting larger anti-ferromagnetic magnetic constant than the CASCI procedure.
This result, although very promising, is not definitive, and in fact correlation effects,
not accounted for in the present work, such as dynamic correlation effects 
outside the active space and convergence with basis set, 
could further enhance the deviation from the biquadratic Heisenberg Hamiltonian.

Considering the results of the present work and the ones available in the literature\cite{Sharma2014, Noodleman1984, Noodleman1992, Chilkuri2019, Neese2020, Presti2019, Gillum1976}(see the SI\cite{SI}for details)
some clear trends can be promptly recognized: increasing the active space,
performing CASSCF orbital optimization and/or recovering dynamic correlation (MCPDFT), 
widens the energy spread of the spin ladder, and, thus, yielding a larger effective 
magnetic coupling coefficient $J$.
The almost doubling of the extracted $J$ and $J'$ due to the CASSCF procedure,
as seen in Figure~\ref{fig:fe2-22in26-spinladder} and~\ref{fig:heisenberg-vs-cas}a, indicates the important 
role of orbital relaxation by differentially stabilizing the low-spin state.

This finding clearly shows that one needs to be cautious when using CI energies on ROHF orbitals,
and a systematic error is to be expected that overstabilizes higher-spin states over low-spin states.
Moreover, the deviation from the simple bilinear Heisenberg model, although small
indicates that the complexity of the interactions in $\ce{[FeS]}$ clusters cannot simply be reduced
to a Heisenberg spin-system when aiming at quantitative accuracy; instead,
more involved forms of interactions are present,
that require complex \emph{ab initio} Hamiltonians (here exemplified by large CASSCF calculations)
and model Hamiltonians (here exemplified by the biquadratic Heisenberg).

{\textbf{CASSCF effects on local spin measurements for Fe$_2$S$_2$}}\\
To further investigate the applicability of a (biquadratic) Heisenberg spin model,
we look into local spin measurements and spin-spin correlation functions between
the two iron centers, and study the CASSCF effect on these quantities.
We explain in Appendix~\ref{sec:local-spin}, how we directly measure these quantities and in App.~\ref{app:spin-corr} and ~\ref{sec:cum-spin-from-rdm} how
to extract them from the spin-free 1- and 2-RDMs. 
We want to emphasize that we are aware that the local spin and spin-spin correlation functions between single and sums of orbitals are representation-dependent quantities. 
Meaning they are not actual physical observables, but they do depend on the type of employed orbitals, i.e. 
localized or delocalized orbitals.
However, they are still an extremely useful means to provide insight in the chemical and physical properties of 
compounds and accordingly, are extensively used in the literature.\cite{Sharma2014, Herrmann2005, RamosCordoba2012, Clark2001}

To ensure reproducibility of our results we want to point out the protocol to obtain the orbitals we used again: 
The starting orbitals for all calculations, were the (10e,10o) ROHF orbitals, for which the iron $\ce{3d}$ and $\ce{3d}'$ 
were identified and separately localized with the default options of the Pipek-Mezey\cite{Pipek1989} method in \texttt{OpenMolcas}\cite{Molcas2016, OpenMolcas2019}. 
These orbitals were then relaxed during the Stochastic-CASSCF procedure and the converged orbitals, which remained very localized and in the initial atom-separated order (discussed further below), were used to obtain the corresponding local spin and spin-spin correlation functions. 
We tested the stability of these results by (a) localizing the final CASSCF orbitals and (b) performing a Procrustes~\cite{Schnemann1966, Weser2021} transformation to map the CASSCF orbitals as close as possible to the starting ROHF orbitals and found no effect on the obtained local spin and spin-spin correlation functions.

\begin{figure*}
	\includegraphics[width=0.8\textwidth]{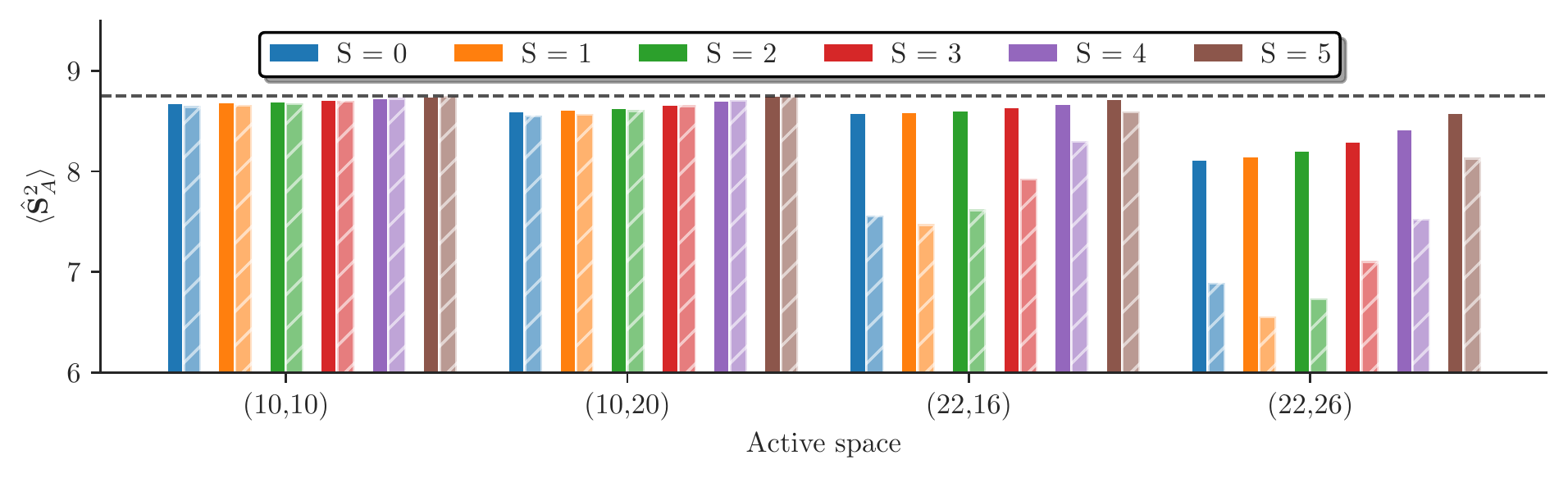}
	\caption{\label{fig:loc-spin-vs-cas}The local spin expectation value on iron $A$, $\braket{\hat{\mathbf{S}}_A^2}$, from the CASCI (solid bars) and the final CASSCF results (striped bars) as a function of the active space size for all spin states. The dashed line indicates the maximal possible value of 8.75.}
\end{figure*}

Figure~\ref{fig:loc-spin-vs-cas} shows the local spin expectation value on
iron $A$, $\braket{\hat{\mathbf{S}}_A^2}$, extracted from the CASCI (solid bars), and the  
CASSCF wave functions (striped bars), for different active spaces and all accessible 
spin states.
The CAS(10e,10o) and CAS(10e,20o) exhibit a local spin expectation value close to the
maximum possible, $\braket{\hat{\mathbf{S}}_A^2}_\mathrm{max} = \frac{5}{2}(\frac{5}{2} + 1) = 8.75$, 
for all spin states. The CASSCF orbital relaxation does not have a significant impact on it.
Upon inclusion of the bridging sulfur orbitals in the CAS(22e,16o)
-- enabling ligand-to-metal (``super-exchange-type'') excitations -- 
the local spin expectation value remains close to the maximum for
CASCI results. However, it substantially drops for all 
spin states upon CASSCF orbital relaxation.
This behavior is enhanced for the CAS(22e,26o), however, for this choice of active space
a reduced local spin expectation value for the low-spin states is already obtained for the CASCI calculations.
Interestingly, the triplet in the CAS(22e,16o) and CAS(22e,26o), and the quintet
in the CAS(22e,26o) have a lower local spin expectation value than the singlet
after the CASSCF procedure.

The CASSCF local spin expectation value of the triplet state in the CAS(22e,26o) 
of $\braket{\hat{\mathbf{S}}_A^2}_\mathrm{min}\!\approx\! 6.5$ corresponds to a 
local spin of $S_A \!\approx\! 2$, which raises the question of the applicability of the Heisenberg model mapping.
In general for systems with local spin momenta larger than $S \!=\! \nicefrac{1}{2}$, local non-Hund
excited states, can cause deviations from a pure Heisenberg behavior.~\cite{Malrieu2013, Anderson1959,
	Bastardis2007, Boca1999, de_Graaf_Broer_book}
Additionally, as discussed by Sharma \emph{et al.}~\cite{Sharma2014} the deviation from 
the pure $S\!=\!\nicefrac{5}{2}$ ion demands accounting for spin and charge delocalization.
Both contributions can be related to additional biquadratic terms in the Heisenberg Hamiltonian.~\cite{Malrieu2008, Anderson1959, Kittel1960, Malrieu2013}

\begin{figure}
	\includegraphics[width=0.3\textwidth]{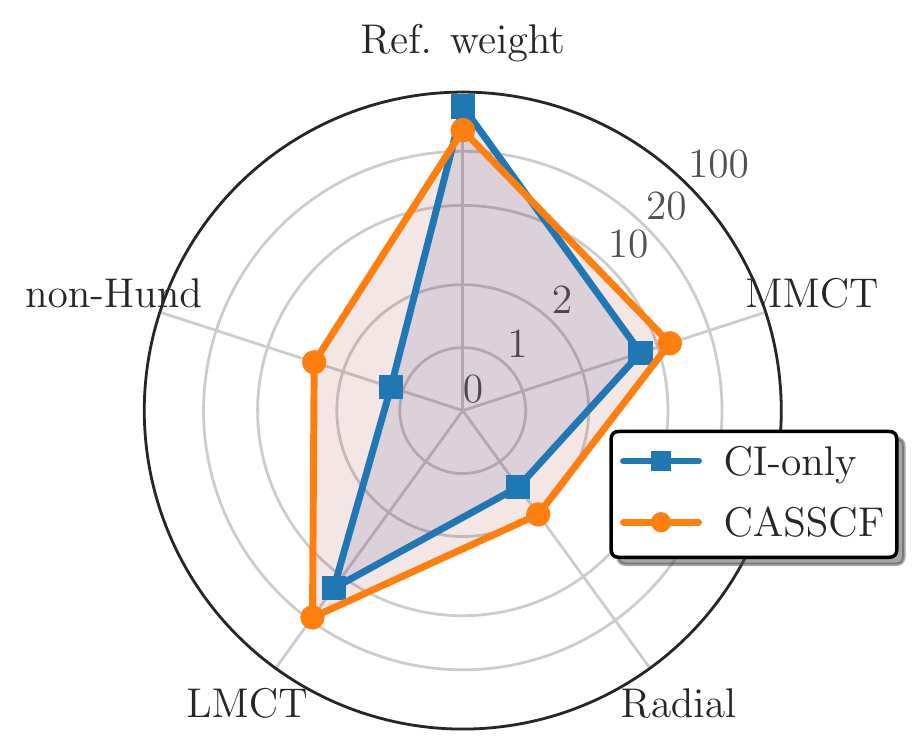}
	\caption{\label{fig:singlet-22in26-wf}Radar plot showing the most important contributions to the CASCI (blue squares) and CASSCF (orange circles) singlet ground state in the (22e,26o) active space in percent. The figure shows the reference weight inter-iron $3d \leftrightarrow 3d$ charge transfer (MMCT), intra-iron ``breathing''-like $3d \ra 3d'$ radial (Radial), bridging-sulfur-to-metal CT (LMCT) and local Hund's rule-violating intra-iron $3d \ra 3d$ excitations (non-Hund).}
\end{figure}

One striking advantage of our methodology, based on the FCIQMC algorithm applied onto localized and site-ordered MOs,
is that we have direct access to the stochastic representation of the ground state wave function. 
Thus, to further analyze the deviations from a pure Heisenberg model, 
we investigated the leading contributions to the CASCI and CASSCF results for each studied active space. 
As an example, for the CASCI (blue squares) and CASSCF (orange circles) singlet results in the (22e,26o) active space, we show the reference weight (Ref. weight), the sum of all metal-to-metal charge transfer (MMCT), local $d \ra d'$ radial excited configurations (Radial), ligand-to-metal charge transfer (LMCT) and 
local Hund's rule violating configurations (non-Hund) in a radar plot 
in Figure~\ref{fig:singlet-22in26-wf}. 
It is important to note, that the values are displayed in percent and the radial axes (indicated by the above introduced acronyms) are on a logarithmic scale to allow an easier visual comparison of the different contributions to the ground state wave function and 
the explicit values can be found in the SI.

The reference weight of the $S = 0$ state in the (22e,26o) active space 
drops from a value of 74.4\% in the CASCI to 46.1\% in the CASSCF wave function. 
On the other hand, the inter-iron MMCT ($Fe_A 3d \leftrightarrow Fe_B 3d$) increase from 6.9\% to 12.9\% and the bridging-sulfur-to-metal LMCT increase 
from an already large 13.4\% to a substantial 27.9\% between the CASCI and
CASSCF calculations. 
Both the radial-type, intra-iron $3d \ra 3d'$, (CASCI: $1.5\%$, CASSCF: $2.1\%$) and intra-iron non-Hund configurations (CASCI: $1.2\%$, CASSCF: $3.7\%$) only have marginal contributions in the wave functions.
The remaining spin states show similarly large LMCT contributions after 
the CASSCF procedure in the (22e,26o) active space.

These results suggest that the main driving force in lowering the 
local spin expectation values are LMCT configurations upon inclusion 
of the bridging-sulfur orbitals in the active space. 
However, as shown in Figure~\ref{fig:loc-spin-vs-cas} by the relatively constant (close-to-maximum) CASCI local spin expectation 
values for all active space, ``just'' including the sulfur 3p orbitals does not suffice to correctly 
capture all relevant correlation mechanisms; instead, 
the CASSCF orbital relaxation of the ROHF starting orbitals is necessary.

Malrieu \emph{et al.},~\cite{Malrieu2002, Calzado2009} Angeli and Calzado~\cite{Calzado2012}
and Li Manni and Alavi~\cite{LiManni2018}
have observed that CASSCF orbitals from a minimal active space are \emph{too localized} to
correctly capture all relevant physical mechanisms in a subsequent second-order multi-reference 
perturbation theory (MRPT2). 
This is mainly due to the fact that relevant ligand-to-metal charge transfer (LMCT) 
excitations do not interact with the zeroth-order wave function due to the
generalized Brillouin theorem.~\cite{Gaston1968, Caballol2002, Calzado2002}
On the other hand, natural magnetic orbitals, obtained by e.g. difference-dedicated CI (DDCI) 
calculations,~\cite{Broer1986, Caballol1992, Malrieu1993, Malrieu1995, Muoz2004} 
or optimized CASSCF orbitals from large active space calculations,~\cite{LiManni2018,LiManni2019,Bogdanov2018}
show \emph{correlation-induced} metal-ligand delocalization, 
by capturing higher-order contributions.~\cite{Calzado2009, Malrieu2002,LiManni2018}

We also studied this effect in the present work by directly comparing the 
localized high-spin $S = 5$ (10e,10o) ROHF orbitals (used as the starting orbitals in all CASSCF calculations) 
with the singlet (22e,26o) CASSCF orbitals. 
During the Stochastic-CASSCF procedure, performed with \texttt{OpenMolcas}, the orbitals 
remain quite localized and in the chosen atom-separated order, mentioned above and described in the SI.
For reproducibility, it is important to note, that we used the last orbitals of the \texttt{OpenMolcas} CASSCF procedure \emph{before} the standard final diagonalization of the 1-RDM and transformation to natural (delocalized) orbitals. 
Furthermore, we performed invariant Procrustes orthogonal transformations~\cite{Schnemann1966, Weser2021} -- with the \texttt{OpenMolcas} software package -- of the (10e,10o) ROHF iron 3d orbitals to make them as similar
as possible to the (22e,26o) singlet CASSCF orbitals, to allow an optimal comparison. 
Further details of the exact protocol for the comparison and corresponding orbital files can be found in the SI.

\begin{figure*}
	\includegraphics[width=\textwidth]{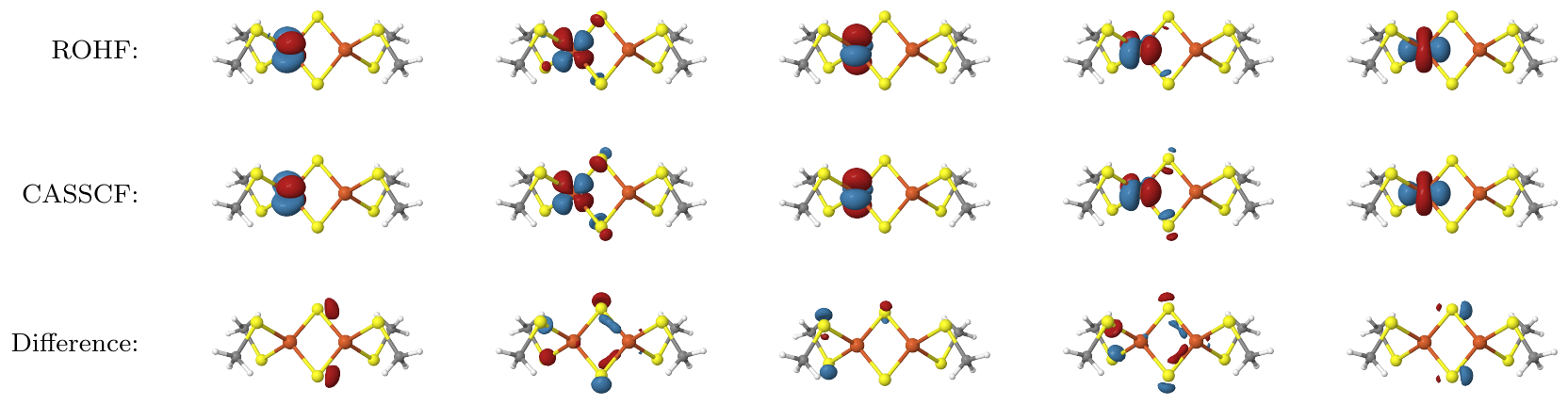}
	\caption{\label{fig:orbial-difference-rohf} (10e,10o) ROHF (top row) and  (22e,26o) $S = 0$ CASSCF Fe$_A$ 3d orbitals rendered with \texttt{Jmol}~\cite{jmol} with an isosurface value of 0.05.
		The difference between the corresponding ROHF and CASSCF orbitals (bottom row) was 
		obtained with \texttt{Multiwfn}~\cite{Lu2011} and is rendered with an 
		isosurface cutoff of 0.007 (except the 3rd column, which uses a value of 0.003).
		The protocol to obtain the orbitals and their differences is described in the main text and with more detail in the SI, where also the corresponding orbital files can be found.
	}
\end{figure*}

In Figure~\ref{fig:orbial-difference-rohf} we show the (10e,10o) ROHF (top row) and 
the CASSCF(22e,26o) singlet (middle row) 3d orbitals of iron $A$, rendered with the
\texttt{Jmol} software package,~\cite{jmol} with an isosurface cutoff value of 0.05.
The last row of Figure~\ref{fig:orbial-difference-rohf} shows the difference of the corresponding orbitals, computed with the \texttt{pegamoid.py}\cite{pegamoid} and \texttt{Multiwfn} software package,~\cite{Lu2011} and rendered with 
\texttt{Jmol} with an isosurface cutoff value of 0.007 for all orbitals except the third (3rd column), which has a cutoff value of 0.003 to make differences visible. 
The delocalization effect of the CASSCF procedure can be seen 
for orbitals two (2nd column) and four (4th column). 
The orbital differences show that the CASSCF procedure has a metal-to-ligand delocalization 
effect, where larger tails of the iron 3d orbitals on the ligands,
increases both the kinetic and direct exchange integrals,~\cite{Anderson1950, Anderson1963} 
and consequently increasing the absolute value of $J$.~\cite{Calzado2009, Calzado2012}
As discussed above and shown in Figure~\ref{fig:singlet-22in26-wf}, the delocalization of the iron 3d orbitals 
is accompanied by a simultaneous increase of the LMCT contributions in the   
(22e,26o) CASSCF singlet wave function.

Calzado \etal,~\cite{Calzado2009} show a very similar orbital dependence when performing CASCI
calculation on extracted $J$ parameters. 
Their study on local $S \!=\! 1$ binuclear systems shows that the
high-spin triplet ROHF orbitals yield a much too low $J$ compared to using singlet or state-specific orbitals.
Similarly, Spiller \etal,~\cite{Neese2020} find that when using spin state-averaged CASSCF orbitals, a subsequent
NEVPT2 treatment yields lower magnetic coupling than using spin-pure state-specific orbitals.
Angeli and Calzado~\cite{Calzado2012} suggest to use average orbitals of the singlet ground
and excited states in the minimal active space
to include the ionic contributions and thus ligand-metal delocalization
and Kubas~\cite{Kubas2020} used spin-averaged Hartree-Fock (SAHF)~\cite{Stavrev1997} orbitals for the low-lying excited state spectrum of the [FeS] dimer.

On the other hand, CASSCF misses different physical effects, which
tends to emphasize the ionic nature of orbitals~\cite{Chilkuri2019} and causes MOs of pure
ionic wave functions to be too diffuse.~\cite{Angeli2009}
Similarly, Malrieu \etal~\cite{Malrieu2002} showed that the definition of magnetic orbitals from spin-unrestricted DFT calculations strongly
overestimate the metal-ligand delocalization, which might be the reason for the
rather large $J$ value obtained by BS-DFT~\cite{Noodleman1984, Noodleman1992} and DMRG CASCI calculations based on such orbitals.~\cite{Sharma2014}

\textbf{CASSCF effect on spin-spin correlation function for Fe$_2$S$_2$}\\
With a spin-adapted basis and the localized and atom ordered 
MOs described in the SI, we can use the formulas derived in 
Appendix~\ref{app:spin-spin} to study the spin-spin interaction
between the two magnetic centers in the Fe$_2$S$_2$ system, and the 
effect of the CASSCF procedure on it. 

\begin{figure*}
	\includegraphics[width=0.8\textwidth]{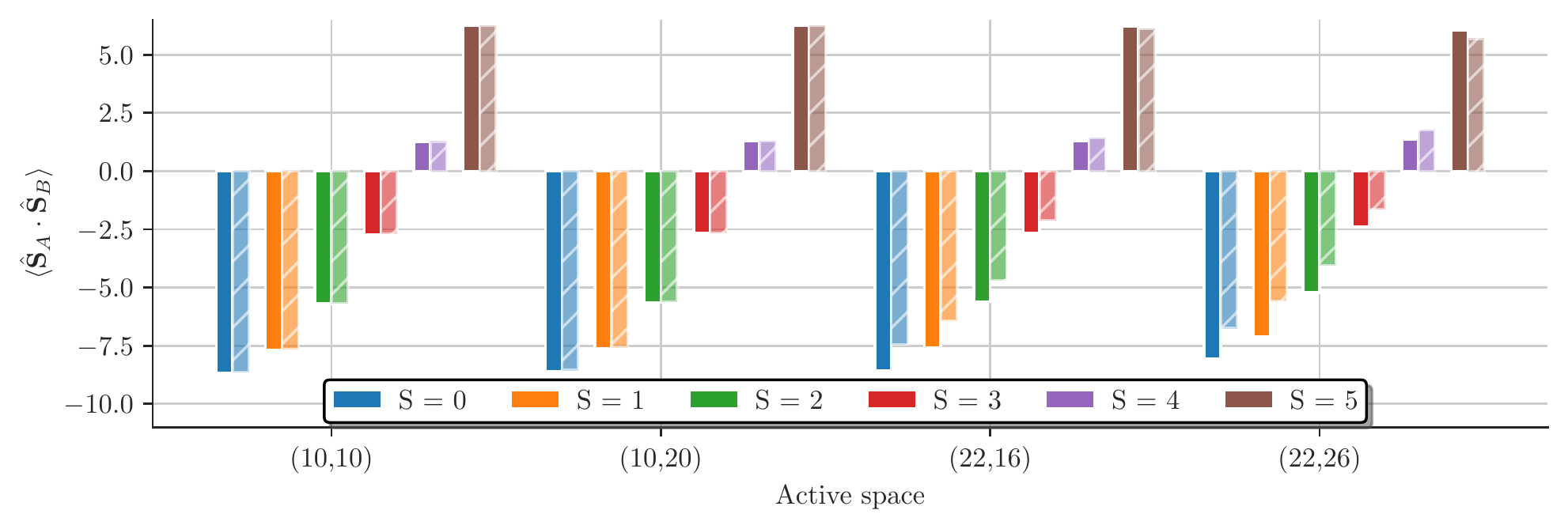}
	\caption{\label{fig:spin-spin-vs-cas}The spin-spin correlation function $\braket{\hat{\mathbf{S}}_A \cdot \hat{\mathbf{S}}_B}$ between local spins on iron $A$ and $B$ from CASCI (solid bars) and the final CASSCF results (striped bars) as a function of the active space size for all spin states.}
\end{figure*}

Figure~\ref{fig:spin-spin-vs-cas} shows the spin-spin correlation
function $\braket{\hat{\mathbf{S}}_A \cdot \hat{\mathbf{S}}_B}$ between 
the two magnetic centers from the CASCI (solid), 
and in the CASSCF wave functions (striped bars), 
as a function of the active space size for all spin states.
For all active spaces the spin-spin alignment changes from anti-ferromagnetic to ferromagnetic,
starting from $S\!=\!4$, as the total spin increases.
The spin-spin correlations are somewhat large for the CAS(10e,10o) and CAS(10e,20o), 
where the CASSCF orbital relaxation does not have a big impact on the expectation values.
As for the local spin measurements, the orbital relaxation has the biggest effect in the CAS(22e,26o). 
The CASSCF procedure has a damping effect on the magnitude of the spin-spin correlations, 
but does not change the description of the underlying physical behavior of a transition 
from an anti-ferromagnetic to a ferromagnetic alignment as a function of the total spin.

\begin{figure*}
	\includegraphics[width=\textwidth]{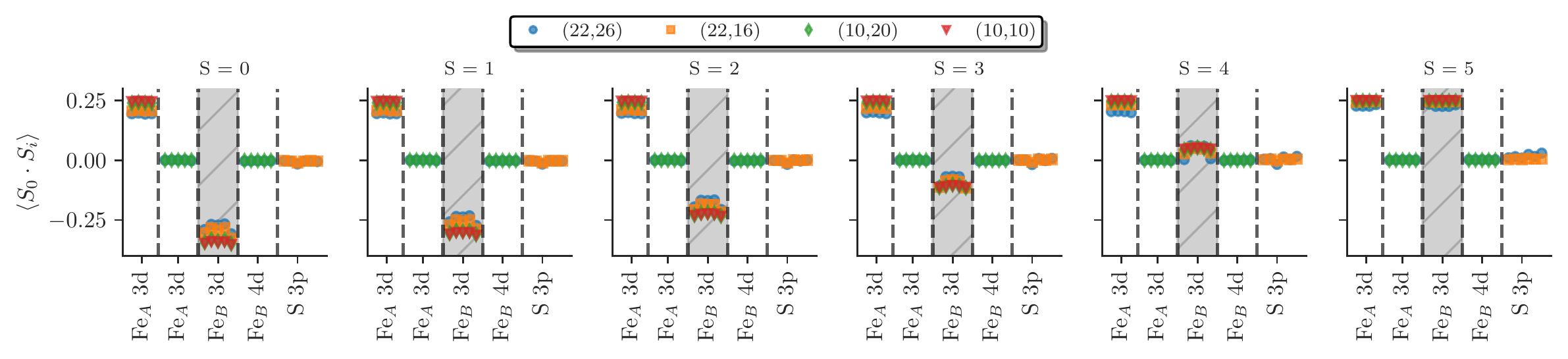}
	\caption{\label{fig:spin-spin-orbitals}Spin-spin correlation function $\braket{\hat{\mbf S}_0 \cdot \hat{\mathbf{S}_i}}$ between the first Fe$_A$ 3d orbital (index 0) and all the other orbitals $i$, obtained via the spin-free RDMs for all the spin states of the  CASSCF results. The x-axis indicates the type of orbitals, where
		the 3d orbitals of iron $B$ are indicated by the gray background. This plot combines all results
		from the different active spaces indicated by the color and marker type (see legend and main text).}
\end{figure*}

With access to the 1- and 2-RDM we are able to study the spin correlation functions on an orbital-resolved
level, including the iron 3d$'$ and sulfur 3p orbitals.
Figure~\ref{fig:spin-spin-orbitals} shows the CASSCF spin-spin correlation function $\braket{\hat{\mbf S}_0 \cdot \hat{\mathbf{S}_i}}$ between the first Fe$_A$ 3d orbital and all other orbitals, obtained via the spin-free RDMs.
Fig.~\ref{fig:spin-spin-orbitals} contains $\braket{\hat{\mbf S}_0 \cdot \hat{\mathbf{S}_i}}$ for all spin states, $S = 0$ to $S = 5$ (indicated by the subplot titles) and all active spaces, different color and markers. 
The x-axes indicate the different orbitals $i$ and different types of orbitals (iron, sulfur, etc.) are separated by vertical dashed lines and data points only show up, when possible. E.g. there are no markers of the (10e,10o) active space results (red triangles) for the iron 4d and sulfur 3p orbitals. 
The mostly singly occupied first iron A 3d orbital, with index 0, is magnetically parallel aligned to all the other Fe$_\text{A}$ 3d orbitals, as can be seen by the $\braket{\hat{\mbf S}_0 \cdot \hat{\mathbf{S}_i}} \approx 0.25, \, \forall i \in \{\text{Fe}_A\, 3d\}$,  all the spin states. 
$\braket{\hat{\mbf S}_0 \cdot \hat{\mathbf{S}_i}} \approx 1/4$ is expected for two ferromagnetically $S = 1/2$ spins.
The magnetic 3d orbitals of iron $B$ are highlighted by the gray background in Fig.~\ref{fig:spin-spin-orbitals}. 
Here on can see that with increasing total spin $S$, indicated by the titles of the subplots, the alignment of 
the first iron $A$ 3d orbital changes from anti-ferromagnetic, $\braket{\hat{\mbf S}_0 \cdot \hat{\mathbf{S}_i}} < 0$, 
to ferromagnetic alignment, $\braket{\hat{\mbf S}_0 \cdot \hat{\mathbf{S}_i}} \approx 1/4, \forall i \in \{\text{Fe}_B\,3d\}$, with the 3d orbitals of iron $B$. 
The results confirm that the exchange interaction exclusively
happens between the (magnetic) iron 3d orbitals, while the other orbitals are magnetically inert (indicated by a zero value of 
$\braket{\hat{\mbf S}_0 \cdot \hat{\mathbf{S}_i}}$).

\subsection{Fe$_4$S$_4$ system \label{sec:fe4}}
We now turn to the all-ferric [Fe(III)$_4$S$_4$(SCH$_3)_4)]$  system.
Here we consider the minimal (20e,20o) active space consisting of
the iron 3d orbitals of the 4 iron atoms.
This active space size is already slightly above the
current limit of performing routine FCI calculations.~\cite{Molcas2016, OpenMolcas2019}
Similar to Fe$_2$S$_2$ we performed state-specific and spin-pure
Stochastic-CASSCF calculations for all the spin states, from $S \!=\! 0$ up to $S \!=\! 10$.
We used the geometry studied in
References~[\citen{LiManni2020b, Sharma2014}],
which is, among other computational details, documented in the SI.
We used an ANO-RCC-VDZ basis set for Fe and an ANO-RCC-MB for all other elements 
and ensured that the obtained results are converged w.r.t. number of used walkers $N_w$.
They are already with a very modest $N_w = 1\cdot 10^6$ walkers.

In  the Fe$_4$S$_4$ study we use the localized (20e,20o) high-spin ROHF orbitals as a
starting guess and focus on the effect of the CASSCF orbital relaxation on the
extraction of the model parameters and associated physical
and chemical interpretations of the results. Similar as in the FeS-dimer section above, we refer to the first iteration of the CASSCF procedure, based on the ROHF orbitals, as CASCI.
In our previous work,~\cite{LiManni2020b} we found that the \emph{ab initio} CASCI results
can be very well mapped to a simple bilinear Heisenberg model.
However, as seen in Section~\ref{sec:fe2} on the dimer model, orbital relaxation effects can 
affect the relative energy of the ab initio spin states, 
and introduce forms of interactions that go beyond the simple bilinear Heisenberg model.

\begin{figure}
	\includegraphics[width=0.35\textwidth]{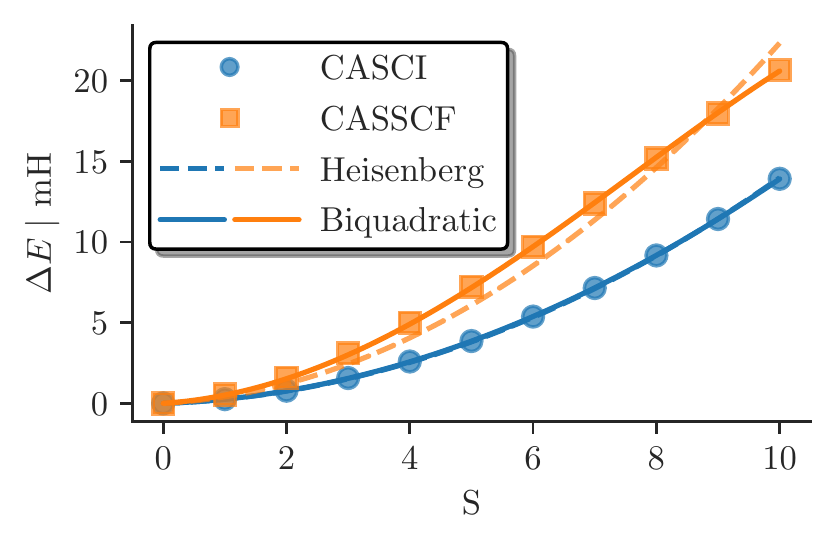}
	\caption{\label{fig:fe4-energies}Energy difference to the $S = 0$ ground state (markers) and a simple (dashed-) and biquadratic (solid lines) Heisenberg fit for the CASCI (blue) and CASSCF (orange) results of the (20e,20o) active space as a function of the total spin.}
\end{figure}

Figure~\ref{fig:fe4-energies} shows the energy difference to the $S \!=\! 0$ ground state (markers) and a simple (solid) and a biquadratic (dashed line) Heisenberg fit for CASCI and CASSCF results as a function of the total spin.
The CASCI results are well represented by a simple Heisenberg model, whereas the
CASSCF results differ from it and necessitate  a biquadratic model description, similar to the
above studied Fe$_2$S$_2$ case.

\begin{figure*}
	\includegraphics[width=0.9\textwidth]{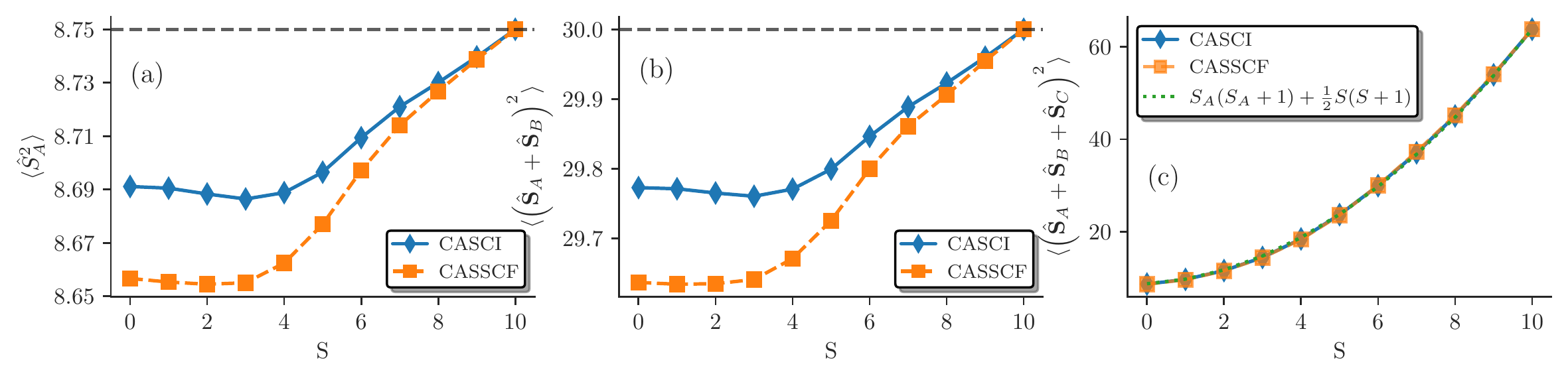}
	\caption{\label{fig:fe4-loc-spin}Local spin expectation values for iron $A$ (a), $A + B$ (b) and $A + B + C$ (c) as a function of total spin for the CASCI (blue) and CASSCF (orange) results.}
\end{figure*}

To investigate the deviation of the \emph{ab initio} CASSCF results from a pure Heisenberg model,
we computed the local spin and spin-spin correlation for the Fe$_4$S$_4$ system.
Figure~\ref{fig:fe4-loc-spin} shows the local spin expectation values of iron $A$ (a), $A + B$ (b) and $A + B + C$ (c) as a function of total spin for the CASCI and CASSCF results. The local spin on the single iron $A$ is close to the maximum possible, $\left( S_{A}^\mathrm{max}\right) ^2 \!=\! 8.75$,
for all the spin states, and the effect of the CASSCF orbital relaxation is present, but small.
Due to symmetry reasons we can safely assume that this expectation value is equal for all 4 iron centers.
The expectation value of the sum of the local spin of the far-distanced irons, $A$ and $B$, is close  to the maximum possible, $\left( S_{AB}^\mathrm{max} \right) ^2 \!=\! 5 (5 +1) \!=\! 30$, as can be seen in Figure~\ref{fig:fe4-loc-spin}b.
This shows that the two far-distanced iron centers are ferromagnetically aligned with $S_A + S_B \!=\! 5$, as already investigated thoroughly for the singlet ground and excited states in
Reference~[\citen{LiManni2020b}]. Again, the orbital relaxation only plays a minor role, but shows
the same behavior as for the single iron spin.
The local spin expectation value of the sum of  three irons, $\braket{( \hat{\mathbf{S}}_A+ \hat{\mathbf{S}}_B+ \hat{\mathbf{S}}_C )^2}$, increases from a minimum value close to
$\left(S_{ABC}^\mathrm{min}\right)^2 \!=\! 8.75$ for $S \!=\! 0$ all the way to $\left(S_{ABC}^\mathrm{max}\right)^2 = \nicefrac{15}{2} (\nicefrac{15}{2} + 1) \!=\! 63.75$, for $S \!=\! 10$.
$\braket{( \hat{\mathbf{S}}_A+ \hat{\mathbf{S}}_B+ \hat{\mathbf{S}}_C )^2}$
can be represented by $S_A (S_A + 1) + \nicefrac{1}{2}S (S + 1)$ as can be seen in the
right panel of Figure~\ref{fig:fe4-loc-spin}, which is the exact results of a 4-site pure $S \!=\! \nicefrac{5}{2}$ Heisenberg model.
The close agreement to the theoretically maximal values of the local spin is because
we investigate the minimal (20e,20o) active space of the magnetic iron 3d orbitals.
Inclusion of ligand orbitals would cause a larger deviation similar to the iron dimer studied above,
and as already anticipated in our previous work.~\cite{LiManni2020b}

\begin{figure*}
	\includegraphics[width=0.9\textwidth]{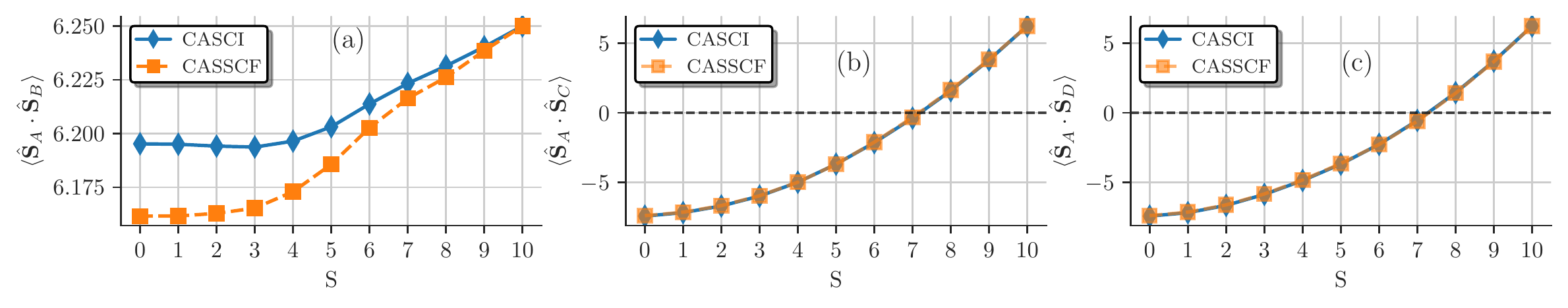}
	\caption{\label{fig:fe4-spin-corr}Spin-spin correlation function, $\braket{\hat{\mathbf{S}}_i \cdot \hat{\mathbf{S}}_j}$ between the 4 different iron atoms for the CASCI (blue) and CASSCF (orange) results as a function of the total spin. (a) shows the spin-spin interaction between the far-distanced magnetic centers A and B and
		(b) and (c) the symmetric interaction between the close-distanced irons, A-C and A-D respectively.}
\end{figure*}

We now focus on the spin-spin interaction between the 4 magnetic iron centers.
Figure~\ref{fig:fe4-spin-corr} shows the  spin-spin correlation function, $\braket{\hat{\mathbf{S}}_i \cdot \hat{\mathbf{S}}_j}$ between the 4 different iron atoms for the CASCI and CASSCF results as a function of the total spin. 
Figure~\ref{fig:fe4-spin-corr}a confirms that the two iron atoms with the largest distance, $A$ and $B$, always stay ferromagnetically aligned
for all the spin states, with a marginally lowering effect of the CASSCF orbital relaxation.
Figure~\ref{fig:fe4-spin-corr}b and c show that the spin-spin interaction between two close lying iron atoms, e.g.~ $A-C$ or $A-D$, is
anti-ferromagnetic for the low spin states, and switches to ferromagnetic alignment for $S = 8$ and higher.
Additionally, these results confirm that these spin-spin interactions are symmetric and that the CASSCF procedure has only a marginal effect on the obtained expectation values.

\section{\label{sec:conclussion}Conclusion}

In this work, we present our implementation to compute the spin-pure one- and two-body reduced density matrices, via stochastic sampling, within our spin-adapted FCIQMC implementation.
This gives us access to spin-pure two-body observables, such as the spin-spin correlation function,
and allows to use the GUGA-FCIQMC as a spin-pure CI eigensolver in the spin-pure
Stochastic-CASSCF approach (within \texttt{OpenMolcas}).
This in turn enables us to stochastically, yet accurately, treat active spaces far larger than
conventional CI solvers in a spin-pure manner.
The implementation requires only minor modification to the existing GUGA-FCIQMC implementation and introduces only a small
computational overhead. This makes the approach quite efficient and allows us to employ up to hundreds of millions 
of CSFs simultaneously.

We demonstrate the utility of this method by studying two FeS dimer and tetramer model systems.
For the dimer, by performing extensive state-specific CASSCF calculations for the lowest
state of each accessible spin-symmetry, and 
four active spaces, we find that:
\textbf{(1)} the combined effect of Fe 3d orbital relaxation and the ligand-to-metal charge transfer has
a larger influence on the energetics of the spin-ladder than the sum of the two effects alone.
\textbf{(2)} When using (10e,10o) ROHF starting orbitals for the CASSCF procedure, its effect is rather small (few mH)
on the singlet-triplet gap, while up to $\approx 20$ mH for low-spin-high-spin gap.
\textbf{(3)} When one maps \emph{ab initio} results to a (biquadratic) Heisenberg Hamiltonian, performing a spin-pure
CASSCF procedure has a large impact on the extracted model parameter.

Access to the spin-pure RDMs with GUGA-FCIQMC, allows us to directly measure local-spin and spin-spin correlation functions.  Insight into these quantities, the local (double) occupation number and the electron
delocalization effect due to the CASSCF procedure, enable us to argue why the CASCI results
using (10e,10o) ROHF orbital agree so well with the bilinear Heisenberg model, while the converged CASSCF do not.
The ROHF orbitals are optimized such that the Heisenberg exchange mechanism, is the only possible one.
Thus, they are too localized on the iron atoms~\cite{Malrieu2002, Calzado2009, Calzado2012} and even increasing the active space
does  not enable to fully capture important spin delocalization and charge fluctuations.~\cite{Noodleman1992, Sharma2014, Malrieu2008}

We study the FeS tetramer in the minimal (20e,20o) active space, which in a spin-adapted approach, due to 20 open shell
localized 3d orbitals  is a formidable task.
Also, for the tetramer we find that performing a CASSCF procedure necessitates in the inclusion of the
biquadratic term into the spin model to correctly map the \emph{ab initio} results.

\section*{Acknowledgements}
The authors thank Thomas Schraivogel (MPI-FKF) for valuable scientific discussions.
The authors gratefully acknowledge financial support by the Max Planck Society.
This project has received funding from the European Union's Horizon 2020 research and innovation programme under Grant Agreement \#952165. The results contained in this paper reflect the authors' view only, and the EU is not responsible for any use that may be made of the information it contains.

\appendix

\section{Local spin measurements \label{sec:local-spin}}

In the GUGA approach, CSFs are spin-eigenfunctions up to any spatial orbital $i$.
This means it is straightforward to calculate the expectation value of a
cumulative local spin operator $\mathbf{\hat S_{c}}(i)$ consisting of orbitals up to the chosen
orbital $i$
\begin{equation}
\label{eq:local-spin}
\mathbf{\hat S_c}(i) = \sum_{j = 1}^i \mathbf{\hat s}_j, \quad \mathbf{\hat s}_j = \left(
\hat s_j^x, \hat s_j^y, \hat s_j^z \right),
\end{equation}
where $\mathbf{\hat s}_j$ indicates the local spin operator of a single molecular orbital (MO) $j$.
The square of the operator defined in Eq.\eqnref{eq:local-spin}, $\mathbf{\hat S_c}^2(i)$, is diagonal in a GUGA-CSF basis, and thus one can straightforwardly, calculate the expectation value
\begin{equation}
\label{eq:loc-spin-exp}
\braopket{\Psi}{\mathbf{\hat S_c}^2(i)}{\Psi} = \sum_\mu 
c_\mu^2\, \braopket{\mu}{\mathbf{\hat S_c}^2(i)}{\mu} = \sum_\mu c_\mu^2\,
S_i^\mu(S_i^\mu + 1),
\end{equation}
where $S_i^\mu$ indicated the intermediate total spin of CSF $\ket{\mu}$ at spatial orbital $i$.
It is important to note, since Eq.\eqnref{eq:loc-spin-exp} is a diagonal quantity, 
the replica method\cite{Alavi2014} needs to be used within FCIQMC to obtain unbiased 
estimates.

Additionally, this necessitates to order orbitals of interest, e.g.~to measure the local spin of a specific iron atom, consecutively starting from the beginning, since GUGA CSFs are not spin-eigenfunctions  for intermediate orbitals.
It turns out that this ordering, in conjunction with using localized 3d$'$ orbitals
in the CAS(22e, 26o) for the Fe$_2$S$_2$ system, is even more optimal as the choice
studied in Reference~\citen{LiManni2020}. More optimal in the sense, that we do have
an even higher reference weight (0.67 compared to 0.55 for the singlet CASCI)\textendash more single reference character\textendash and hence a faster convergence.
The detailed orbital choice and ordering used can be found in the SI.

\section{Spin-spin interaction \label{app:spin-spin}}
The measurement of local spin quantities additionally allows us to compute the
spin-spin interaction between different iron sites.

\textbf{2 sites}

If we assume a set of local, independent spin operators $\hat S_i$, with $\commutator{\hat S_i}{\hat S_j} = 0$ and due to symmetry:
$\braket{\hat S_i^2} = \braket{\hat S_j^2} \forall \;i, j$, we can deduce the following relations:
\begin{align}
\label{eq:2-sites}
\braket{( \hat S_A + \hat  S_B)^2} &= \braket{ \hat S_A^2 + \hat S_B ^2 + \hat S_A \cdot \hat S_B + \hat S_B \cdot \hat S_A}  \nonumber \\
&= 2 \left( \braket{\hat S_A^2} +  \braket{\hat S_A \cdot \hat S_B} \right) \nonumber \\
\Rightarrow \braket{\hat S_A \cdot \hat S_B} &= \frac{1}{2}  \braket{( \hat S_A + \hat  S_B)^2}  -  \braket{\hat S_A^2}.
\end{align}
Since, following Eq.\eqnref{eq:loc-spin-exp}, we can measure both $  \braket{( \hat S_A + \hat  S_B )^2}$ and $ \braket{\hat S_A^2}$ locally, we can deduce the spin correlation function from purely local spin measurements.

\textbf{3 sites}

\begin{figure}
	\includegraphics[width=0.3\textwidth]{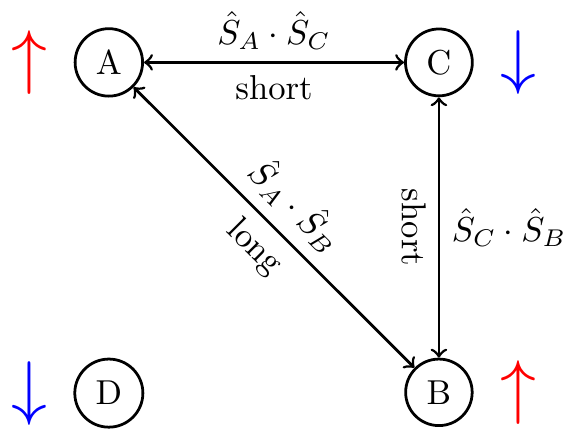}
	\caption{\label{fig:geometry} Sketch of the Fe$_4$S$_4$ geometry.}
\end{figure}

Next, we consider a model system as the one depicted in Fig.~\ref{fig:geometry} with two long bond distances $AB$ and $CD$ and 4 remaining short distances and assume $\braket{\hat S_i^2}$ to be identical for all $i$. 
If we assume spin correlations functions to be equal for
the same bond distances, e.g. $\braket{\hat S_A \cdot \hat S_B} = \braket{\hat S_C \cdot \hat S_D}$ and $\braket{\hat S_A \cdot S_C} = \braket{\hat S_B \cdot \hat S_C}$ etc, we can deduce, with $ \hat S_A + \hat S_B + \hat S_C = \hat S_{ABC}$ for short, 
{\small 
	\begin{align}
	\label{eq:3-sites}
	\braket{(S_{ABC} )^2} &= 3 \braket{\hat S_A^2} + 2 ( \braket{\hat S_A \cdot \hat S_B} + \braket{\hat S_A \cdot \hat S_C} + \braket{\hat S_B \cdot \hat S_C} ) \nonumber \\
	&=  3 \braket{\hat S_A^2}  + 2 \braket{\hat S_A \cdot \hat S_B} + 4 \braket{\hat S_A \cdot \hat S_C} \nonumber \\
	&= 3\braket {\hat S_A^2}  +  \braket{( \hat S_A + \hat  S_B )^2} - 2\braket{\hat S_A^2} +  4 \braket{\hat S_A \cdot \hat S_C} \nonumber \\
	\Rightarrow  \braket{\hat S_A \cdot \hat S_C} &= \frac{\braket{( \hat S_{ABC} )^2} - \braket{( \hat S_A + \hat  S_B )^2} - \braket {\hat S_A^2} }{4},
	\end{align}
}
where again we can measure all quantities on the right of Eq.~(\ref{eq:3-sites}) directly, via Eq.\eqnref{eq:loc-spin-exp}.

\textbf{4-sites}\\
Now we assume: all $\braket{\hat S_i^2}$ are the same, $\commutator{\hat S_i}{\hat S_j} = 0, \forall\; i, j$, $\braket{\hat S_A \cdot \hat S_B} = \braket{\hat S_C \cdot \hat S_D}$, $\braket{\hat S_A \cdot \hat S_C} = \braket{\hat S_B \cdot \hat S_C}$ and $\braket{\hat S_A \cdot \hat S_D} = \braket{\hat S_B \cdot \hat S_D}$.
With $ \hat S_A + \hat S_B + \hat S_C  + \hat S_D = \hat S_{tot}$ for short, we obtain 
\begin{align}
\label{eq:4-sites-2}
\braket{\hat S_{tot}^2} =& 4\braket{\hat S_A^2} + 2\Big(\braket{\hat S_A \cdot \hat S_B} + \braket{\hat S_A \cdot \hat S_C} + \braket{\hat S_A \cdot \hat S_D} \nonumber \\
&+ \braket{\hat S_B \cdot \hat S_C} + \braket{\hat S_B\cdot \hat S_D} + \braket{\hat S_C \cdot \hat S_D} \Big) \nonumber \\
=& 4 ( \braket{\hat S_A^2} + \braket{\hat S_A \cdot \hat S_B} + \braket{\hat S_A \cdot \hat S_C} + \braket{\hat S_A \cdot \hat S_D} )
\end{align}
And plugging Eq.~(\ref{eq:2-sites}) and  (\ref{eq:3-sites}) into Eq.~(\ref{eq:4-sites-2}) yields
\begin{align}
\braket{\hat S_A \cdot \hat S_D} =& \frac{1}{4}\braket{\hat S_{tot}^2}  -  \braket{\hat S_A^2} -\left( \frac{1}{2}\braket{S_{AB}^2} - \braket{\hat S_A^2}\right) \nonumber \\ 
&-\frac{1}{4} \left[ \braket{\hat S_{ABC}^2} - \braket{\hat S_{AB}^2} - \braket{\hat S_A^2} \right]  \nonumber \\
=& \frac{1}{4} \braket{\hat S_{tot}^2} - \frac{1}{2}\braket{\hat S_{AB}^2} + \frac{1}{4} \braket{\hat S_{AB}^2} \nonumber \\
&- \frac{1}{4} (\braket{\hat S_{ABC}} - \braket{\hat S_A^2} )\nonumber \\
=& \frac{1}{4}\left(\braket{\hat S_{tot}^2} - \braket{\hat S_{ABC}^2} - \braket{\hat S_{AB}^2} + \braket{S_A^2}\right),
\end{align}

with $\hat S_{AB} = \hat S_A + \hat S_B$ and $\hat S_{ABC} = \hat S_A + \hat S_B + \hat S_C$.

\section{\label{app:spin-corr}Orbital resolved local spin and spin correlation function from spin-free RDMs}

Expressing the local spin operators as~\cite{Paldus2012}
\begin{equation}\label{eq:local-spin-op}
S_i^k = \sum_{\mu,\nu = \uparrow,\downarrow} = \sigma_{\mu,\nu}^{k} a_{i,\mu}^\dagger a_{i\nu},
\end{equation}
with the Pauli matrices~\cite{Pauli1925}
\begin{equation}\label{eq:pauli-matrices}
\sigma^x =
\begin{pmatrix}
0	&	\phantom{-}1 \\	1 & \phantom{-}0 \\
\end{pmatrix}, \quad
\sigma^y =
\begin{pmatrix}
0 & -i \\ i & \phantom{-}0 \\
\end{pmatrix}, \quad
\sigma^z =
\begin{pmatrix}
1 & \phantom{-}0 \\ 0 & -1 \\
\end{pmatrix}
\end{equation}
and the fermionic creation (annihilation) operators, $a_{i,\s}^{(\dagger)}$ of electrons with spin $\s$ in spatial orbital $i$.
This results in the explicit expressions
\begin{align}
S_i^x &= \frac{1}{2}\left( a_{i\u}^\dagger a_{i\d} + a_{i\d}^\dagger a_{i\u} \right) \\
S_i^y &= \frac{1}{2}\left( a_{i\d}^\dagger a_{i\u} - a_{i\u}^\dagger a_{i\d} \right) \\
S_i^z &= \frac{1}{2}\left( n_{i\u} - n_{i\d} \right),
\end{align}
where $n_{i\s} = a_{i\s}^\dagger a_{i\s}$ is the fermionic number operator of orbital $i$ and spin $\s$.

If we express the $\hat{\mathbf{S}}_i \cdot \hat{\mathbf{S}}_j$ as
\begin{equation}\label{eq:spin-corr-start}
\hat{\mathbf{S}}_i \cdot \hat{\mathbf{S}}_j = \hat{{S}}_i^z \cdot \hat{{S}}_j^z + \hat{{S}}_i^x \cdot \hat{{S}}_j^x + \hat{{S}}_i^y \cdot \hat{{S}}_j^y
\end{equation}
and consequently the individual terms as
\begin{align}
\hat{{S}}_i^z \cdot \hat{{S}}_j^z =& \frac{1}{4}\left(n_{i\u} - n_{i\d}  \right)\left(n_{j\u} - n_{j\d}  \right) \\
\hat{{S}}_i^x \cdot \hat{{S}}_j^x =& \frac{1}{4}\Big( \cre{i\u} \ann{i\d} \cre{j\u} \ann{j\d}  + \cre{i\u} \ann{i\d} \cre{j\d} \ann{j\u} \nonumber \\
&+\cre{i\d} \ann{i\u}  \cre{j\u} \ann{j\d} + \cre{i\d} \ann{i\u} \cre{j\d} \ann{j\u} \Big) \\
\hat{{S}}_i^y \cdot \hat{{S}}_j^y =& \frac{1}{4}\Big(\cre{i\d} \ann{i\u} \cre{j\d} \ann{j\u} -
\cre{i\d} \ann{i\u} \cre{j\u} \ann{j\d} \nonumber \\
&- \cre{i\u} \ann{i\d} \cre{j\d} \ann{j\u} +
\cre{i\u} \ann{i\d} \cre{j\u} \ann{j\d}  \Big).
\end{align}

Combining the $x$ and $y$ terms yields
\begin{equation}\label{eq:x-y-sum}
\hat S_i^x \cdot \hat S_j^x + \hat S_i^y \cdot \hat S_j^y
= \frac{1}{2}\sum_{\s} \cre{i\s} \ann{i\bar{\s}} \cre{j\bar{\s}} \ann{j\s}.
\end{equation}

\underline{For $i = j$} we can transform \eqnref{eq:x-y-sum} to
\begin{align}
\hat S_i^x \cdot \hat S_j^x + \hat S_i^y \cdot \hat S_j^y =&
\frac{1}{2}\sum_{\s} \cre{i\s} \ann{i\bar{\s}} \cre{j\bar{\s}} \ann{j\s} \nonumber \\
=& \frac{1}{2}\sum_\s \num{i\s}(1 - \num{i\bar\s})
\end{align}

For the total local spin operator this means
\begin{equation}
\hat{\mathbf{S}}_i \cdot \hat{\mathbf{S}}_i = \hat{S}^{z^2}_i + \frac{1}{2}\sum_\s \num{i\s}(1 - \num{i\bar\s}),
\end{equation}
and consequently we get the relation
\begin{align}\label{eq:on-site-spin-corr}
S_i^2 &= S_i^{z^2} + \frac{1}{2} \sum_\s \num{i\s} - \frac{1}{2} \sum_\s \num{i\s} \num{i\bar{\s}} \nonumber \\
S_i^2 - \frac{1}{2}\sum_\s \num{i\s} &= \frac{1}{4}(\num{i\u} - \num{i\d})^2 - \frac{1}{2}\sum_\s \num{i\s} \num{i\bar{\s}}\nonumber \\
&= \frac{1}{4} (\num{i\u}^2 - 2\num{i\u}\num{i\d} + \num{i\d}^2 ) - \num{i\u}\num{i\d} \nonumber \\
S_i^2 &= \frac{3}{4}\left(\sum_\s \num{i\s} - \sum_\s \num{i\s}\num{i\bar{\s}} \right).
\end{align}
With the spin-free excitation operators $\hat E_{ij} = \sum_\s \cre{i\s}\ann{i\s}$, and $\hat e_{ij,kl} = \hat E_{ij}\hat E_{kl} - \delta_{jk}\hat E_{il}$, we can express \eqnref{eq:on-site-spin-corr} simply as
\begin{equation}\label{eq:loc-spin-uga}
S_i^2 = \frac{3}{4}\left(\hat E_{ii} - \hat e_{ii,ii}\right),
\end{equation}
since $E_{ii} = \sum_\s \num{i\s}$ and $\hat e_{ii,ii} = \hat E_{ii}\hat E_{ii} - \hat E_{ii} = (\num{i\u} + \num{i\d})^2 - \hat E_{ii} = 2\num{i\u}\num{i\d}$.

With the spin-free excitation operators we can write \eqref{eq:on-site-spin-corr} as
\begin{equation}\label{eq:start-spin-as-uga}
S_i^2 = S_i^{z^2} + \frac{1}{2}\left(\hat E_{ii} - \hat e_{ii,ii}\right)
\end{equation}
and substituting \eqref{eq:loc-spin-uga} on the lhs of \eqref{eq:start-spin-as-uga}, we get
\begin{equation}
\frac{3}{4}\left( \hat E_{ii} - \hat e_{ii,ii}\right) = S_i^{z^2} + \frac{1}{2}\left( \hat E_{ii} - \hat e_{ii,ii}\right)
\end{equation}
leading to the relation
\begin{equation}
S_i^{z^2} = \frac{1}{4}\left( \hat E_{ii} - \hat e_{ii,ii} \right),
\end{equation}
allowing us to formulate the apparent spin-dependent quantity $S_i^{z^2}$ entirely in spin-free terms for $i = j$.

\underline{For $i\neq j$} we can transform \eqnref{eq:x-y-sum} to
\begin{equation}
\hat S_i^x \cdot \hat S_j^x + \hat S_i^y \cdot \hat S_j^y =  -\frac{1}{2}\sum_\s \cre{i\s} \ann{j\s} \cre{j\bar{\s}} \ann{i\bar{\s}} = \hat A_{ij}.
\end{equation}
With the observation
\begin{align}
\hat E_{ij}\hat E_{ji} =& \left(\sum_\s \cre{i\s}\ann{i\s} \right) \left(\sum_\tau \cre{j\tau}\ann{j\tau} \right) \nonumber \\
=& \underbrace{\sum_\s \cre{i\s}\ann{j\s}\cre{j\bar{\s}}\ann{i\bar{\s}}}_{-2\hat A_{ij}} +
\underbrace{\sum_{\s} \cre{i\s}\ann{j\s} \cre{j\s}\ann{i\s}}_{\num{i\s}(1 - \num{j\s})} \nonumber\\
=&-2 \hat A_{ij} + \sum_\s \num{i\s} - \sum_\s \num{i\s}\num{j\s}  \nonumber \\
=& -2\hat A_{ij} + \hat E_{ii} - \sum_\s \num{i\s}\num{j\s},
\end{align}
leading to the relation
\begin{align}
\hat A_{ij} =& -\frac{1}{2}\left( \hat E_{ij}\hat E_{ji} - \hat E_{ii} + \sum_\s \num{i\s}\num{j\s}\right) \nonumber\\
=& -\frac{1}{2} \left(e_{ij,ji} + \sum_\s \num{i\s}\num{j\s}\right),
\end{align}
with which the spin-correlation, \eqnref{eq:spin-corr-start}, can be expressed as
\begin{equation}\label{eq:spin-corre-inter}
\hat{\mathbf{S}}_i \cdot \hat{\mathbf{S}}_j  = S_i^z \cdot S_j^z - \frac{1}{2} \left( \hat e_{ij,ji} + \sum_\s \num{i\s}\num{j\s}\right).
\end{equation}
To express \eqref{eq:spin-corre-inter} in spin-free terms we can rewrite
\begin{align}
S_i^z \cdot S_j^z - \frac{1}{2}\sum_\s \num{i\s}\num{j\s} =& \frac{1}{4}(\num{i\u} - \num{i\d})(\num{j\u} - \num{j\d}) \nonumber \\
&-\frac{1}{2}(\num{i\u}\num{j\u} + \num{i\d}\num{j\d}) \nonumber \\
= \frac{1}{4}(\num{i\d}\num{j\u} &- \num{i\u}\num{j\d} - \num{i\d}\num{j\u} + \num{i\d}\num{j\d})\nonumber \\
&- \frac{1}{2}(\num{i\u}\num{j\u} + \num{i\d}\num{j\d}) \nonumber \\
=& - \frac{\hat e_{ii,jj}}{4},
\end{align}
which allows us to write the spin-spin correlation function entirely in spin-free terms as
\begin{equation}\label{eq:spin-free-spin-corr}
\hat{\mathbf{S}}_i \cdot \hat{\mathbf{S}}_j  = -\frac{1}{2}\left( \hat e_{ij,ji} + \frac{\hat e_{ii,jj}}{2} \right).
\end{equation}
Eq.\eqnref{eq:spin-free-spin-corr} allows us to directly obtain the off-diagonal $i\neq j$, spin-spin correlation functions $\braket{\hat{\mathbf{S}}_i \cdot \hat{\mathbf{S}}_j}$ from the spin-free 2-RDM elements, $\braket{\hat e_{ij,ji}}$ and $\braket{\hat e_{ii,jj}}$ on an orbital resolved level.

\section{\label{sec:cum-spin-from-rdm}Local spin and spin correlation functions of a sum of orbitals from spin-free RDMs}

The results from the previous two sections, Sec.~\ref{sec:local-spin} and Sec.~\ref{app:spin-corr}, can be combined to obtain the local spin, 
$\mathbf{\hat S_c}^2(i)$ (Eq.\eqnref{eq:local-spin}), and spin-spin correlation function of a sum of orbitals, e.g. between magnetic centers as in the iron $A$ and $B$ 3d orbitals, directly 
from the orbital-resolved spin-free RDMs, $\rho_{ij}$ and $\Gamma_{ij, kl}$, respectively $\braket{\hat E_{ij}}$ and $\braket{\hat e_{ij,kl}}$.

\textbf{Local spin:}\\
To obtain the local spin of set of orbitals $\mathcal{I}$
we need to combine Eqs.\eqnref{eq:loc-spin-uga} and~\eqref{eq:spin-free-spin-corr} to get
\begin{align}\label{eq:sum-loc-spin}
\braket{(\sum_{i\in \mathcal{I}} \mathbf{\hat S}_i)^2} =& \sum_{i\in\mathcal{I}} \braket{\mathbf{\hat S}_i^2} + 
\sum_{i\neq j \in \mathcal{I}} \braket{\mathbf{\hat S}_i \cdot  \mathbf{\hat S}_j} \nonumber\\
=& \frac{3}{4} \sum_{i\in\mathcal{I}} \braket{\hat E_{ii}} - \braket{\hat e_{ii,ii}} \nonumber \\
&-\frac{1}{2}\sum_{i\neq j\in\mathcal{I}} \braket{\hat e_{ij,ji}} + \frac{\braket{\hat e_{ii,jj}}}{2}.
\end{align}

\textbf{Spin-spin correlation function:}\\
Similarly, to obtain the spin-spin correlation function between two sets of orbitals $\mathcal{I}$ and $\mathcal{J}$, we need to make use of Eqs.\eqnref{eq:loc-spin-uga} and~\eqref{eq:spin-free-spin-corr}
\begin{align}\label{eq:sum-spin-corr}
\braket{\sum_{i\in\mathcal{I}}  \mathbf{\hat S}_i \cdot \sum_{j\in\mathcal{J}}  \mathbf{\hat S}_j} &= \sum_{i\in \mathcal{I},j\in\mathcal{J}} \braket{\mathbf{\hat S}_i \cdot \mathbf{\hat S}_j} \nonumber\\
&=  -\frac{1}{2}\sum_{i\in \mathcal{I},j\in\mathcal{J}} \braket{\hat e_{ij,ji}}
+ \frac{\braket{\hat e_{ii,jj}}}{2},
\end{align}
assuming $\mathcal{I} \cap \mathcal J = \emptyset$. If the two sets, $\mathcal{I}$ and $\mathcal{J}$ do overlap, Eq.\eqnref{eq:sum-spin-corr} has to be adapted to use Eq.\eqnref{eq:loc-spin-uga} in the case $i = j$. 

The advantage of Eqs.\eqnref{eq:sum-loc-spin} and \eqnref{eq:sum-spin-corr}, compared to the 
cumulative local-spin and the spin-correlation functions derived in App.~\ref{app:spin-spin}, 
is that they are independent of the ordering of orbitals and do not assume any symmetries of 
the problem at hand. 

\bibliographystyle{aipnum4-1}


%

\end{document}